\documentclass[12pt, draftclsnofoot, onecolumn]{IEEEtran}

\usepackage[T1]{fontenc}
\usepackage{graphicx}
\usepackage{hyperref}
\usepackage{array}
\usepackage{amsmath}
\usepackage{booktabs}
\usepackage{setspace}
\usepackage{float}
\usepackage{xcolor}

\ifCLASSINFOpdf
\else
\fi
\usepackage{amsmath}
\usepackage[cmintegrals]{newtxmath}

\begin{document}
%
\title{Real-Time Radio Technology and Modulation Classification via an LSTM Auto-Encoder}
%
%
%

\author{
Ziqi Ke and Haris Vikalo\\
Department of Electrical and Computer Engineering\\
The University of Texas at Austin\\
ziqike@utexas.edu, hvikalo@ece.utexas.edu
}

\maketitle

\begin{abstract}
Identification of the type of communication technology and/or modulation scheme based on detected radio signal are challenging 
problems encountered in a variety of applications including spectrum allocation and radio interference mitigation. They are rendered
difficult due to a growing number of emitter types and varied effects of real-world channels upon the radio signal. Existing
spectrum monitoring techniques are capable of acquiring massive amounts of radio and real-time spectrum data using compact 
sensors deployed in a variety of settings. However, state-of-the-art methods that use such data to classify emitter types and detect 
communication schemes struggle to achieve required levels of accuracy at a computational efficiency that would allow their 
implementation on low-cost computational platforms. In this paper, we present a learning framework based on an LSTM denoising 
auto-encoder designed to automatically extract stable and robust features from noisy radio signals, and infer modulation or 
technology type using the learned features. The algorithm utilizes a compact neural network architecture readily implemented on 
a low-cost computational platform while exceeding state-of-the-art accuracy. Results on realistic synthetic as well as over-the-air 
radio data demonstrate that the proposed framework reliably and efficiently classifies received radio signals, often demonstrating superior performance compared to state-of-the-art methods.
\end{abstract}


%
\IEEEpeerreviewmaketitle

\section{Introduction}
Analysis of detected radio signals enables classification of communication technology and modulation schemes employed
by the source that emitted the signals; this information helps optimize spectrum allocation and mitigate radio interference, 
supports wireless environment analysis and enables improvement of communication efficiency. However, increase in the 
numbers of emitter types and sources of interference, as well as temporal variations in the effects of wireless environment
on the transmitted signals, render the accurate inference of communication schemes and emitter types computationally 
challenging. 


Existing methods for modulation and technology classification can be organized into two sub-groups, likelihood-based and 
feature-based \cite{Dobre2007}. Likelihood-based methods make a decision by evaluating a likelihood function of the 
received signal and comparing the likelihood ratio with a pre-defined threshold. Although the likelihood-based classifiers 
are optimal in that they minimize the probability of false classification, they suffer from high computational complexity 
\cite{Dobre2007}. On the other hand, feature-based approaches are relatively simple to implement and may achieve near-optimal 
performance but the features and decision criteria need to be carefully designed. Such methods rely on expert features including 
cyclic moments \cite{Gardner1988} and their variations \cite{Gardner1992}, and spectral correlation functions of analog and digital 
modulated signals \cite{Gardner19871,Gardner19872}; \cite{Yu2006} describes novel decision criteria which utilize pre-existing 
expert features. \cite{Fehske2005} facilitates classification via a multilayer perceptron that relies on spectral correlation functions. 
Expert systems have been shown to achieve high accuracy on certain special tasks but may be challenging to apply in general
settings since the crafted features may not fully reflect all the real-world channel effects. As an alternative, deep learning based 
methods that learn directly from the received signals have recently been proposed. In particular, \cite{Shea2016} utilizes a 
convolutional neural network (CNN) that operates on in-phase and quadrature-phase (IQ) data and outperforms expert features based methods. \cite{West2017} 
combines CNN and long short-term memory (LSTM) \cite{Hochreiter1997,Gers1999} to further improve classification accuracy. 
\cite{Rajendran2018} utilizes an LSTM on amplitude and phase data by simply transferring IQ data for modulation classification, 
outperforming the proposed model in \cite{West2017}. \cite{Oshea2018} proposes two classification models with one adapting 
the Visual Geometry Group (VGG) architecture  \cite{VGG} principles to a 1D CNN and the other utilizing the ideas of deep residual networks 
(RNs)  \cite{RN}. Note that while spectrum monitoring devices are capable of acquiring 
detailed wireless signal's IQ components, storage of such data on distributed sensing devices 
or their transmission to a cloud or edge device for processing is often infeasible due to resource constraints. To this end, distributed 
spectrum monitoring systems such as Electrosense \cite{SRajendran2018} formulate technology detection task as the classification 
that uses more compact Power Spectral Density (PSD) data as features. Note, however, that the aforementioned deep learning
architectures are infeasible for use in distributed settings and on low-cost computational platforms. More details on practical aspects
of RF acquisition can be found in \cite{Xu2019}.

In this paper, we propose a new learning framework for both modulation as well as technology classification problems
based on an LSTM denoising auto-encoder. The framework aims to estimate posterior probabilities of the modulation or technology 
types using time domain amplitude and phase of a radio signal. Auto-encoders in an unsupervised manner 
learn a low-dimensional representation of data; more specifically, they attempt to perform a dimensionality reduction while robustly 
capturing essential content of high-dimensional data \cite{goodfellow2016deep}. Typically, auto-encoders consist of two blocks: an 
encoder and a decoder. The encoder converts input data into the so-called codes while the decoder reconstructs the input from the 
codes. The act of copying the input data to the output would be of little interest without an important additional constraint – namely, 
the constraint that the dimension of codes is smaller than the dimension of the input. This enables auto-encoders to extract salient 
features of the input data. A denoising auto-encoder (DAE) \cite{Vincent2008} can help extract stable and robust features by 
introduction noise corruption to the input signal. In our proposed framework, the received radio signals are first partially corrupted and 
the framework then recovers the destroyed signals, simultaneously learning stable and robust low-dimensional signal representations 
and classifying the signals based on the learned features.

Our main contributions are summarized as follows:
\begin{itemize}
\item We propose a new learning framework which uses amplitude and phase data for modulation classification; the framework is 
based on an LSTM denoising auto-encoder and achieves state-of-the-art modulation classification accuracy. 
\item We extend the proposed framework to technology classification using power spectral density data. 
\item The proposed framework achieves significantly higher top-1 classification accuracy while having much simpler structure than
the existing models. This enables real-time modulation and/or technology classification on compact and affordable computational 
devices, as we demonstrate using Raspberry PI platforms.
\end{itemize}

\section{Methods}
\subsection{Problem Formulation}
Let $r_{t_i} = (x_1, ..., x_n)$ denote a sequence of $m$-dimensional features characterizing $n$ samples of the received radio
signal sampled starting at time $t_i$. The goal of modulation (technology) classification is to identify the modulation (technology) type of the radio signal among 
$K$-classes by estimating $P(y = C_k | r_{t_i})$ where $C_k$ denotes the $k^{th}$ class and $y$ is the true class of the
signal. 

\begin{figure*}[!ht]
	\centering
	\includegraphics[width= 1 \columnwidth]{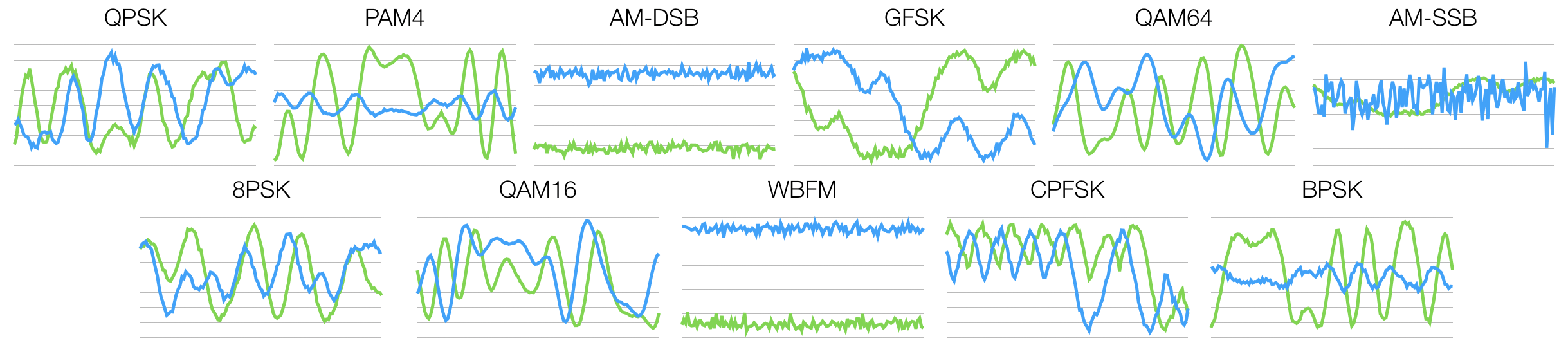}
	\caption{IQ component examples of 11 different modulation types from RadioML2016.10A data with $SNR = 10$dB. 
	The x-axis represents time and the y-axis represents amplitude. The blue line shows the in-phase component while 
	the green line shows the quadrature component.}
	\label{fig:iq}
\end{figure*}

For modulation classification, the features are the IQ components of the sampled signal (i.e., $m = 2$). 
Figure~\ref{fig:iq} shows examples of the IQ components for 11 different modulation types found in
RadioML2016.10A dataset for signal-to-noise ratio (SNR) of $10$dB. Although there are differences 
between the IQ components, it is challenging even for a domain expert to distinguish between them 
due to pulse shaping, distortion and other channel effects \cite{Shea2016}.
\begin{figure*}[!ht]
	\centering
	\includegraphics[width= 1 \columnwidth]{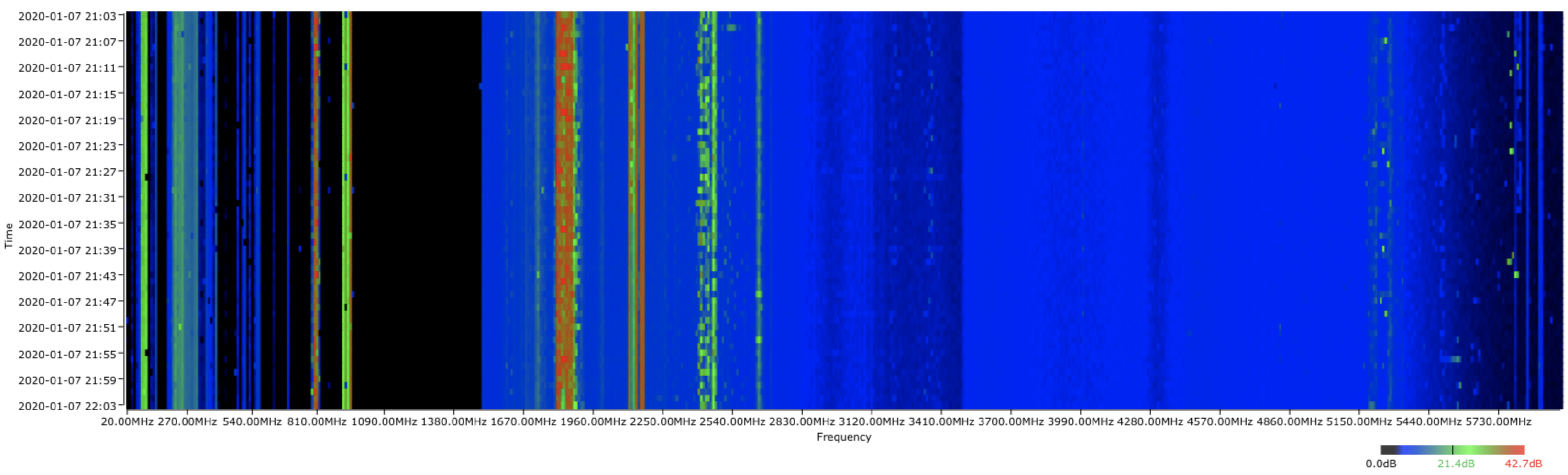}
	\caption{The wireless magnitude spectrum collected by one of the Electrosense sensors. The x-axis contains different 
	centre frequencies while the y-axis shows sampling times.}
	\label{fig:psd}
\end{figure*}

For technology classification, the spectrum of interest is scanned by selecting a candidate carrier frequency $f_j$ in discrete 
increments, and for each such frequency a fast Fourier transform (FFT) of the received signal demodulated into baseband 
is computed. Average values of the FFT coefficients computed for each $f_j$ are then concatenated to form a sequence of
features used to perform the classification task. For the Electrosense data that we analyze in this paper, 
$f_j \in (50\text{MHz}, 6\text{GHz})$ and the scanning resolution is $0.1$MHz.
Figure~\ref{fig:psd} shows an example of wireless magnitude spectrum data from one of the Electrosense sensors.

To characterize the performance of modulation and technology classification methods, we rely on top-1 classification accuracy 
over SNR, confusion matrix, time and space complexity in terms of the number of trainable parameters and model size, and testing time on Raspberry Pi.

\subsection{An LSTM Denoising Auto-Encoder}
In this section, we describe the design of our proposed classifier based on a denoising auto-encoder and recurrent neural 
networks. Instead of using IQ components, for modulation classification we rely on L2-normalized amplitude and normalized phase (falling between -1 and 1, in radians); such normalization benefits learning temporal dependencies \cite{Rajendran2018}. 
\begin{figure}[!ht]
	\centering
	\includegraphics[width= 0.9\columnwidth]{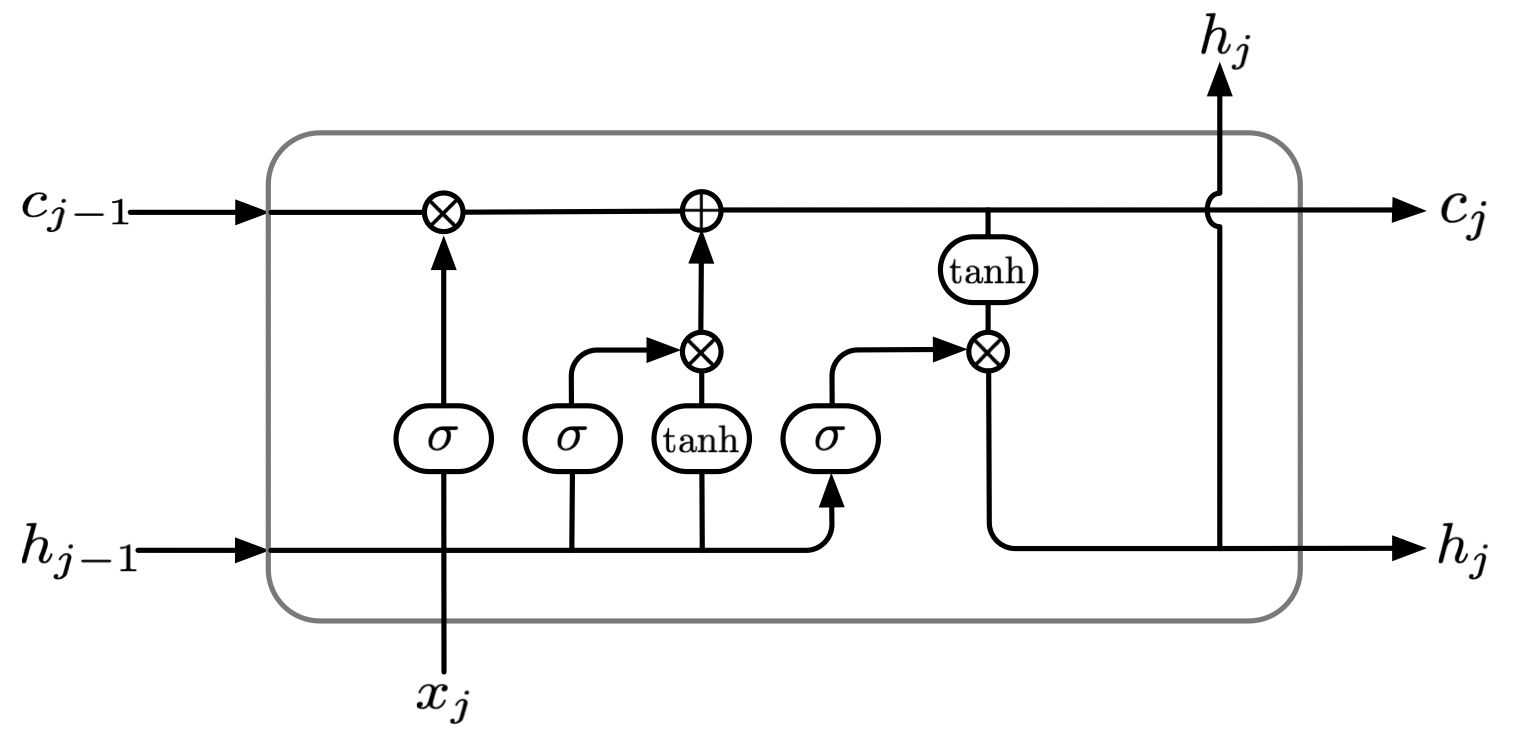}
	\caption{Structure of an LSTM cell.}
	\label{fig:LSTM}
\end{figure}
A sampled radio signal results in a time series, and an LSTM is utilized to efficiently capture temporal structure of such a
series. Figure~\ref{fig:LSTM} shows the structure of an LSTM cell with a forget gate. The input gate, output gate and forget
 gate can be expressed respectively as 
\begin{equation}
i_j = \sigma(W_ix_j + U_ih_{j-1} + b_i),
\end{equation}
\begin{equation}
o_j = \sigma(W_ox_j + U_oh_{j-1} + b_o),
\end{equation}
and
\begin{equation}
f_j = \sigma(W_fx_j + U_fh_{j-1} + b_f),
\end{equation}
while the cell state vector and hidden state vector are defined as
\begin{equation}
c_j = f_t \cdot c_{j-1} + i_j \cdot \tanh(W_cx_j + U_ch_{j-1} + b_c)
\end{equation}
and
\begin{equation}
h_j = o_j \cdot \tanh(c_j),
\end{equation}
respectively, where $\sigma$ denotes the sigmoid function (i.e., $\sigma(z) = \frac{1}{1 + e^{-z}}$), 
$W$ denotes a weight matrix for the input time series, $U$ is a weight matrix for the hidden state 
vector, and $b$ represents a bias vector.

The denoising auto-encoder corrupts the signal by randomly setting a portion of samples of $r(t)$ 
to $0$, thus obtaining a partially destroyed signal $\tilde{r}_{(t)} = (\tilde{x}_1, ..., \tilde{x}_n)$. The partially destroyed
signal is fed to the auto-encoder for training while the original signal is utilized for testing.

Motivated by \cite{Vincent2008}, \cite{chen2018} and \cite{ke2020}, we propose a novel LSTM denoising auto-encoder
for modulation/technology classification where the auto-encoder and classifier are trained simultaneously. 
\begin{figure*}[!ht]
	\centering
	\includegraphics[width= 0.9\columnwidth]{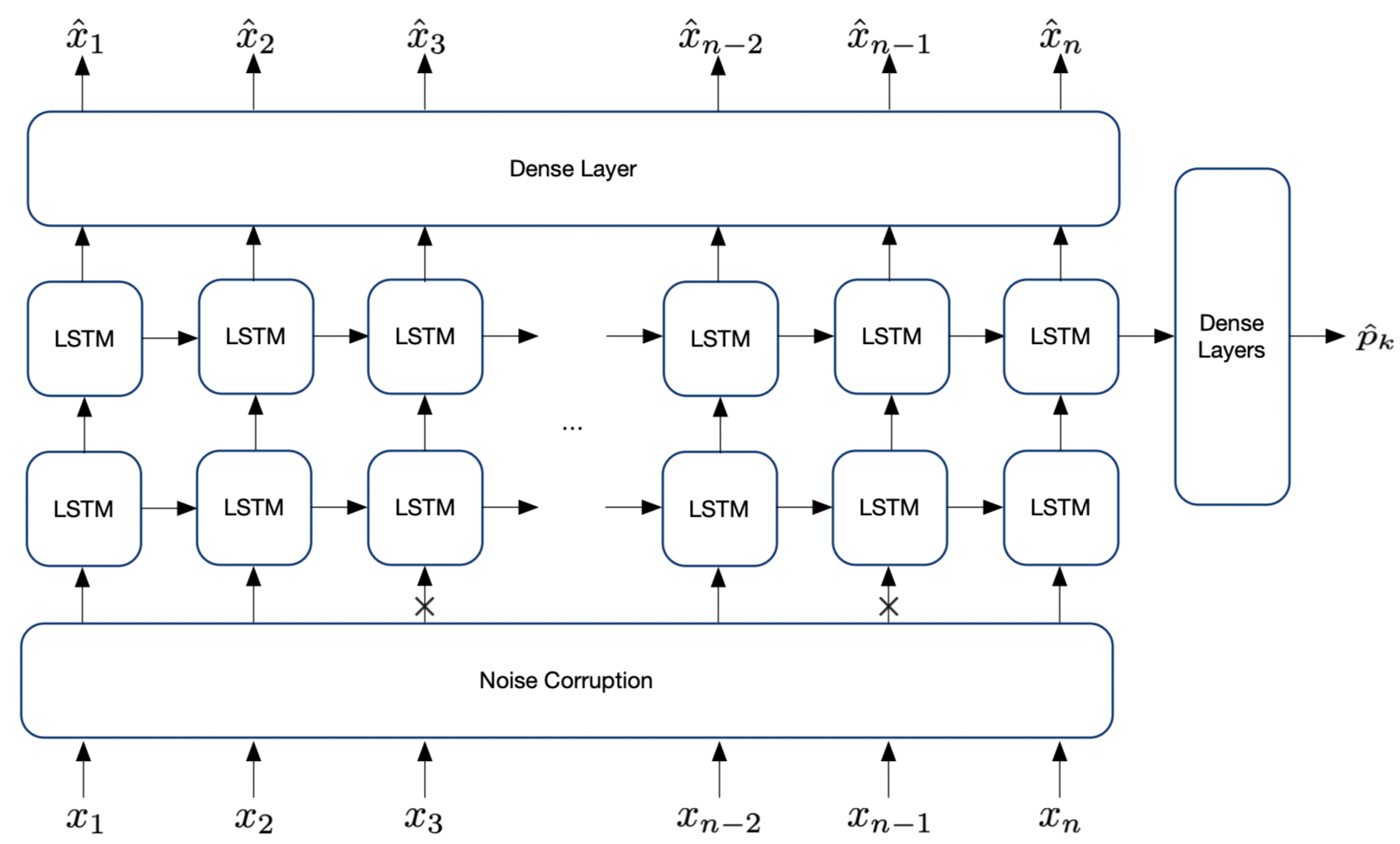}
	\caption{An LSTM denoising auto-encoder classifier.}
	\label{fig:LSTMDAECLF}
\end{figure*}
Figure~\ref{fig:LSTMDAECLF} shows the LSTM denoising auto-encoder classifier. The classifier is 
connected to the last hidden state vector $h_n$ and it consists of 3 fully connected layers followed 
by a softmax function, i.e.,
\begin{equation} 
o^{\text{clf}}_{(1)} = \sigma(W^{\text{clf}}_{(1)}h_n + b^{\text{clf}}_{(1)}),
\end{equation}
\begin{equation} 
o^{\text{clf}}_{(2)} = \sigma(W^{\text{clf}}_{(2)}o^{\text{clf}}_{(1)} + b^{\text{clf}}_{(2)}),
\end{equation}
\begin{equation} 
o^{\text{clf}}_{(3)} = \sigma(W^{\text{clf}}_{(3)}o^{\text{clf}}_{(2)} + b^{\text{clf}}_{(3)}),
\end{equation}
\begin{equation} 
\hat{p}_k = \frac{e^{o^{\text{clf}}_{(3)_k}}}{\sum_{k=1}^Ke^{o^{\text{clf}}_{(3)_k}}},
\end{equation}
where $W^{\text{clf}}$ denotes the weight matrix, $b^{\text{clf}}$ is the bias vector, 
$o^{\text{clf}}$ denotes the output of a fully connected layer for classification
($o^{\text{clf}}_{(3)_k}$ represents its $k^{th}$ entry), $\hat{p}_k$ denotes the 
probability of predicting $r(t)$ as the $k^{th}$ class and $\sigma$ denotes the 
rectified linear unit (ReLU). The 2-layer LSTM operates as an encoder that converts 
the corrupted input $\tilde{r}_{(t)}$ into hidden state vectors $(h_1, ..., h_n)$ while 
a shared fully connected layer operate as a decoder, i.e.,
\begin{equation}
\hat{x}_j = W^{\text{dec}}h_j + b^{\text{dec}},
\end{equation}
where $\hat{x}_j$ is the $j^{th}$ recovered sample, $W^{\text{dec}}$ denotes the 
weight matrix of the decoder and $b^{\text{dec}}$ is the bias vector for the fully 
connected layer. Note that we break the
symmetry of the architecture by using a shared fully connected layer for the
decoder since doing so reduces computational complexity.

Therefore, the loss function of the network consists of the reconstruction loss, $L_\text{DAE}$, and the classification 
loss, $L_{\text{clf}}$. The final loss function is a weighted combination of these
two terms, i.e.,
\begin{equation}
L = (1 - \lambda)L_\text{DAE} + \lambda L_{\text{clf}},
\end{equation}
where $\lambda \in [0, 1]$ is a hyperparameter balancing $L_\text{DAE}$ and 
$L_{\text{clf}}$. It is worth pointing out that a small $\lambda$ eliminates the 
effects of classification layers while a large $\lambda$ distorts the learned 
representation of data. We set the value of $\lambda$ to 0.1 to promote extraction 
of reliable low-dimensional representations of the original signals and thus enable
efficient classification with reduced dimensionality of the hidden state of an LSTM 
cell. This allows the proposed model to achieve higher classification accuracy at a
significantly reduced computational complexity. The reconstruction loss $L_\text{DAE}$ 
is defined to be the mean-squared error (MSE) and can be expressed as
\begin{equation}
L_\text{DAE} = \frac{1}{n}\sum_{j = 1}^n (x_j - \hat{x}_j)^2,
\end{equation}
while the classification loss $L_{\text{clf}}$ is defined to be the categorical 
cross entropy
\begin{equation}
L_{\text{clf}} = -\sum_{k = 1}^K p_k\log \hat{p}_k,
\end{equation}
where $p_k = 1$ if $r(t)$ belongs to the $k^{th}$ class and $p_k = 0$ otherwise.

\subsection{Model Parameters}
For both tasks, we rely on Adam optimizer \cite{Adam2014} since it helps avoid 
local optima. The dimensionality of the hidden states of the LSTM in our denoising
autoencoder is set to $32$. Please note that prior LSTM-based methods require more than 128 
hidden states to achieve desired level of accuracy; otherwise, the classification 
accuracy deteriorates significantly as shown in \cite{Rajendran2018}. The number 
of nodes in the dense layer of decoder is set to $m$ and the number of nodes in the fully connected layers of the classifier 
are set to $32$, $16$ and $K$, respectively. The learning rate is set to $0.001$ and 
the number of epochs is set to $150$. Dropout rate is chosen to be $0.2$ for the
LSTMs and fully connected layers; randomly selected 10\% of the entries of the 
input signal in the training data are masked by $0$. The models are implemented on a computer with
3.70GHz Intel i7-8700K processor, 2 NVIDIA GeForce GTX 1080Ti computer graphics 
cards and 32GB of RAM.  The minibatch size of 128 is utilized. The parameter 
$\lambda$ controlling how the reconstruction loss and classification loss are combined
is set to $0.1$.

\section{Results}
\subsection{Performance Comparison on RadioML2016.10A}
We first evaluate performance of the proposed model for modulation classification on a realistic RadioML2016.10A 
dataset. RadioML data\footnote{\url{https://www.deepsig.io/datasets}} includes a series of synthetic and over-the-air 
modulation classification sets created by DeepSig Inc. Among them, RadioML2016.10A has in particular been
widely used for benchmark testing \cite{Tomothy2016,West2017,Rajendran2018}. Radio channel effects including
time-delay, time-scaling, phase rotation, frequency offset and additive thermal noise are accounted for to emulate 
practical radio communications (details can be found in \cite{OShea2016}). The set contains data for 11 modulation 
schemes (8PSK, AM-DSB, AM-SSB, BPSK, CPFSK, GFSK, PAM4, QAM16, QAM64, QPSK and WBFM). The SNR 
ranges from -20dB to 18dB with 2dB step size; there are $1000$ samples for each SNR resulting in the total of 
220k samples. The sample length is 128. We used 50\%, 25\% and 25\% of the dataset for training, validation and 
testing, respectively. 
\begin{figure}[!ht]
	\centering
	\includegraphics[width= 0.9\columnwidth]{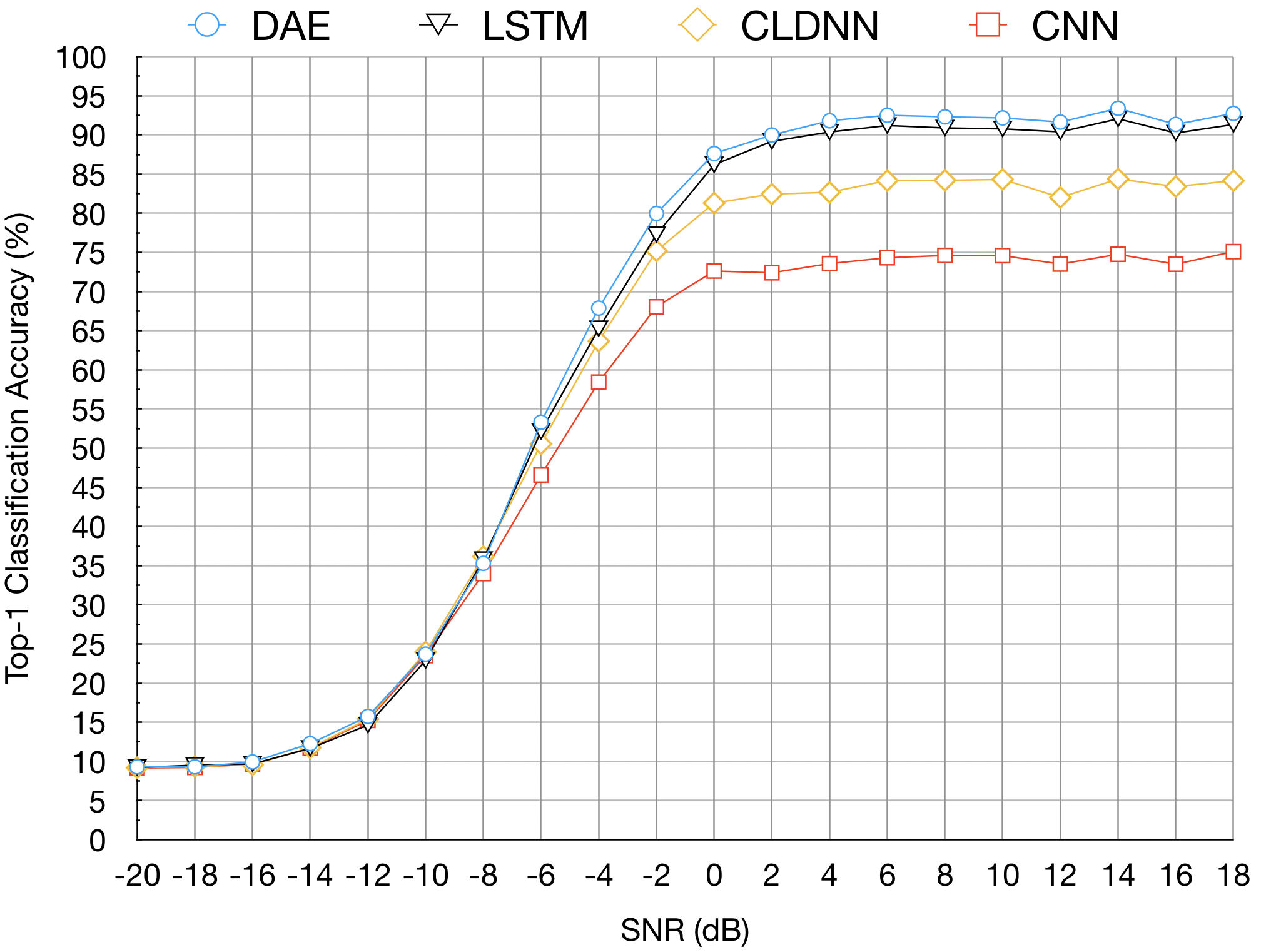}
	\caption{Top-1 classification accuracy of our proposed model (DAE) vs. existing models on 
	RadioML2016.10A dataset.}
	\label{fig:SNR}
\end{figure}

\begin{table*}[!ht]
\caption{Top-1 classification accuracy comparison of our proposed model (DAE) vs. existing models on RadioML2016.10A 
dataset. The highest average top-1 classification accuracy for each SNR is marked in bold.}
\centering
\resizebox{0.95\columnwidth}{!}{\begin{tabular}{>{\centering\arraybackslash}p{0.1\columnwidth} >{\centering\arraybackslash}p{0.1\columnwidth} >{\centering\arraybackslash}p{0.09\columnwidth}>{\centering\arraybackslash}p{0.09\columnwidth} >{\centering\arraybackslash}p{0.09\columnwidth} >{\centering\arraybackslash}p{0.09\columnwidth} >{\centering\arraybackslash}p{0.09\columnwidth} >{\centering\arraybackslash}p{0.09\columnwidth} >{\centering\arraybackslash}p{0.09\columnwidth}}
\toprule
Model  &  & -20dB & -18dB & -16dB & -14dB & -12dB & -10dB & -8dB\\
\midrule
DAE & Mean &\textbf{9.30} & 9.35 & \textbf{9.92} & \textbf{12.27} & \textbf{15.75} & 23.70 & 35.29 \\
& Std & 0.09  & 0.17 & 0.43 & 0.98 & 1.32 & 1.25 & 1.13\\
LSTM & Mean & 9.22  & \textbf{9.51}& 9.67 & 11.66  & 14.64 & 22.89 & 35.78\\
& Std & 0.21 & 0.23  & 0.30 & 0.62 & 1.11 & 1.31 & 1.31\\
CLDNN & Mean & 9.19 & 9.35 & 9.56 & 11.76 & 15.39 & \textbf{23.98} & \textbf{36.16}\\
& Std & 0.16 & 0.22 & 0.23 & 0.74 & 1.17 & 1.38 & 1.61\\
CNN & Mean & 9.15  & 9.23  & 9.64  & 11.68 & 15.24 & 23.51 & 33.98 \\
& Std & 0.13 & 0.26 & 0.40 & 0.79 & 0.98 & 1.31 & 0.82\\

\bottomrule
\end{tabular}}

\vspace{5pt}

\resizebox{0.95\columnwidth}{!}{\begin{tabular}{>{\centering\arraybackslash}p{0.1\columnwidth} >{\centering\arraybackslash}p{0.1\columnwidth} >{\centering\arraybackslash}p{0.09\columnwidth}>{\centering\arraybackslash}p{0.09\columnwidth} >{\centering\arraybackslash}p{0.09\columnwidth} >{\centering\arraybackslash}p{0.09\columnwidth} >{\centering\arraybackslash}p{0.09\columnwidth} >{\centering\arraybackslash}p{0.09\columnwidth} >{\centering\arraybackslash}p{0.09\columnwidth}}
\toprule
Model  &  & -6dB & -4dB & -2dB & 0dB & 2dB & 4dB & 6dB \\
\midrule
DAE & Mean & \textbf{53.31} & \textbf{67.88} & \textbf{79.97} & \textbf{87.62} & \textbf{89.98} & \textbf{91.81} & \textbf{92.61}\\
& Std & 1.56 & 1.23 & 1.08 & 0.74 & 0.60 & 0.45 & 0.23  \\
LSTM & Mean & 52.14  & 65.26 & 77.29 & 86.20 & 89.18 & 90.37 & 91.20\\
& Std & 0.94 & 1.07 & 0.65 & 0.85 & 0.64 & 0.81 & 0.52 \\

CLDNN & Mean & 50.54 & 63.66 & 75.20 & 81.30& 82.45 & 82.68 & 84.18 \\
& Std & 1.30 & 1.57 & 1.53 & 0.83& 0.91 & 1.43 & 0.96 \\
CNN & Mean & 46.56 & 58.43 & 68.05 & 72.62& 72.40 & 73.57 & 74.32\\
& Std & 1.06 & 0.92 & 0.65 & 1.14& 0.91 & 1.56 & 1.15 \\

\bottomrule
\end{tabular}}

\vspace{5pt}

\resizebox{0.95\columnwidth}{!}{\begin{tabular}{>{\centering\arraybackslash}p{0.1\columnwidth} >{\centering\arraybackslash}p{0.1\columnwidth} >{\centering\arraybackslash}p{0.09\columnwidth}>{\centering\arraybackslash}p{0.09\columnwidth} >{\centering\arraybackslash}p{0.09\columnwidth} >{\centering\arraybackslash}p{0.09\columnwidth} >{\centering\arraybackslash}p{0.09\columnwidth} >{\centering\arraybackslash}p{0.09\columnwidth} >{\centering\arraybackslash}p{0.09\columnwidth}}
\toprule
Model  &  & 8dB & 10dB & 12dB & 14dB & 16dB & 18dB & Overall \\
\midrule
DAE & Mean & \textbf{92.31} & \textbf{92.17} & \textbf{91.64} & \textbf{93.40} & \textbf{91.34} &\textbf{ 92.75} & \textbf{61.72} \\
& Std & 0.31 & 0.34 & 0.46 & 0.31 & 0.17 &0.31  & 0.18 \\
LSTM & Mean & 90.89 & 90.77 & 90.39 & 92.05 & 90.24 & 91.33 & 60.49 \\
& Std & 0.50 & 0.58 & 0.78 & 0.63 & 0.69 & 0.78  & 0.37 \\

CLDNN & Mean & 84.21 & 84.32 & 82.02 & 84.38 & 83.41 & 84.16 & 56.78\\
& Std & 1.14 & 0.82 & 0.90 & 0.79 & 1.02 & 0.99 & 0.46\\
CNN & Mean & 74.61 & 74.59 & 73.51 & 74.76 & 73.49 & 75.10 & 51.29 \\
& Std & 0.97 & 1.31 & 1.33 & 1.29 & 1.60 & 1.54 & 0.40 \\

\bottomrule
\end{tabular}}
\label{tab:10a_result}
\end{table*}

\begin{figure}[H]
	\centering
	\includegraphics[width= 0.65\columnwidth]{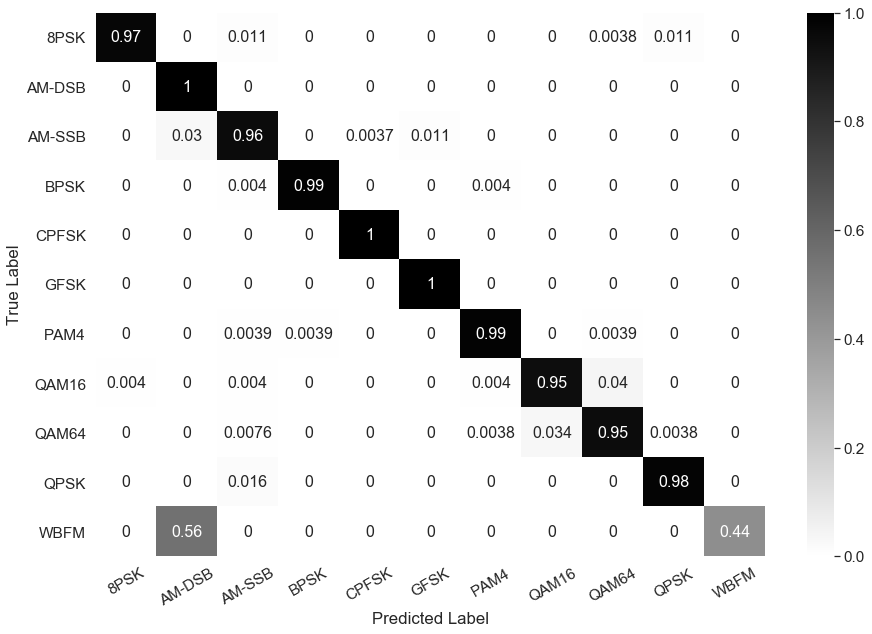}
	\includegraphics[width= 0.65\columnwidth]{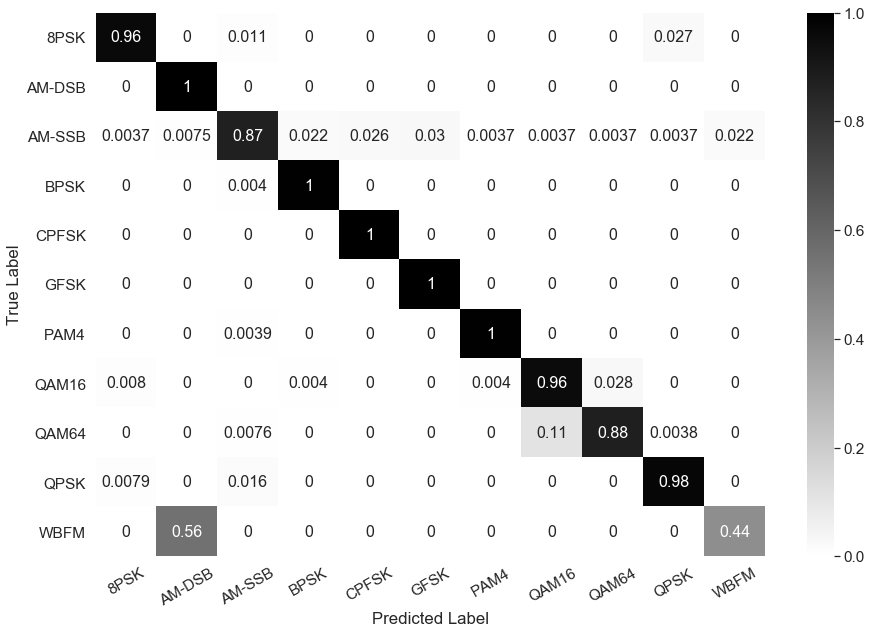}
	\caption{Confusion matrix for the proposed model (top) and LSTM (bottom) on RadioML2016.10A dataset at 18dB SNR.}
	\label{fig:confusion_1}
\end{figure}

\begin{figure}[H]
	\centering
	\includegraphics[width= 0.65\columnwidth]{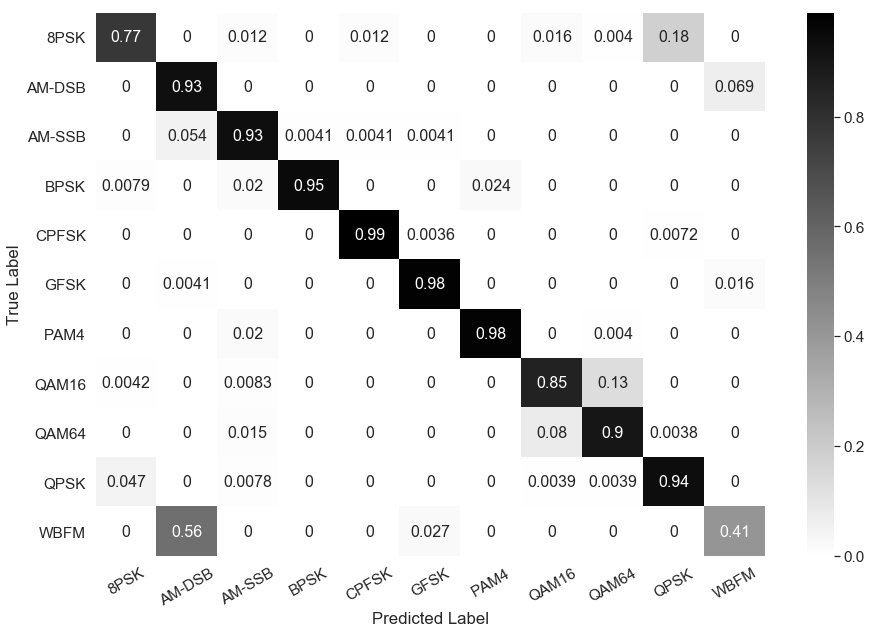}
	\includegraphics[width= 0.65\columnwidth]{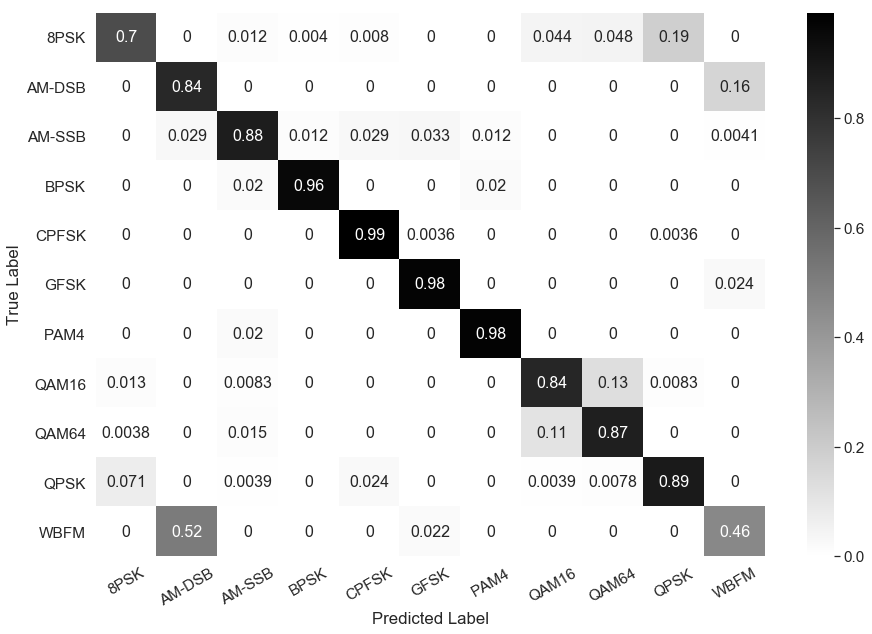}
	\caption{Confusion matrix for the proposed model (top) and LSTM (bottom) on RadioML2016.10A dataset at 0dB SNR.}
	\label{fig:confusion_2}
\end{figure}

\begin{figure}[H]
	\centering
	\includegraphics[width= 0.65\columnwidth]{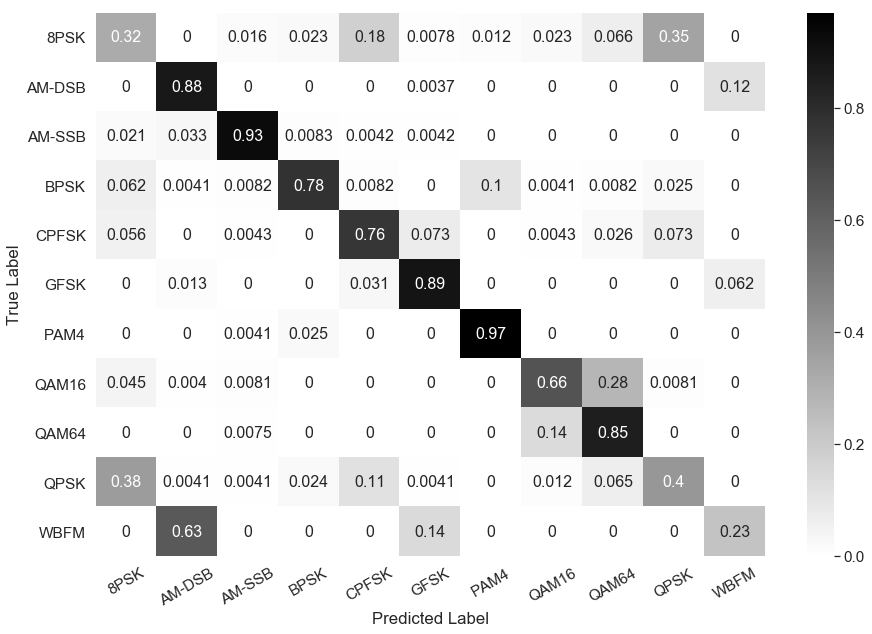}
	\includegraphics[width= 0.65\columnwidth]{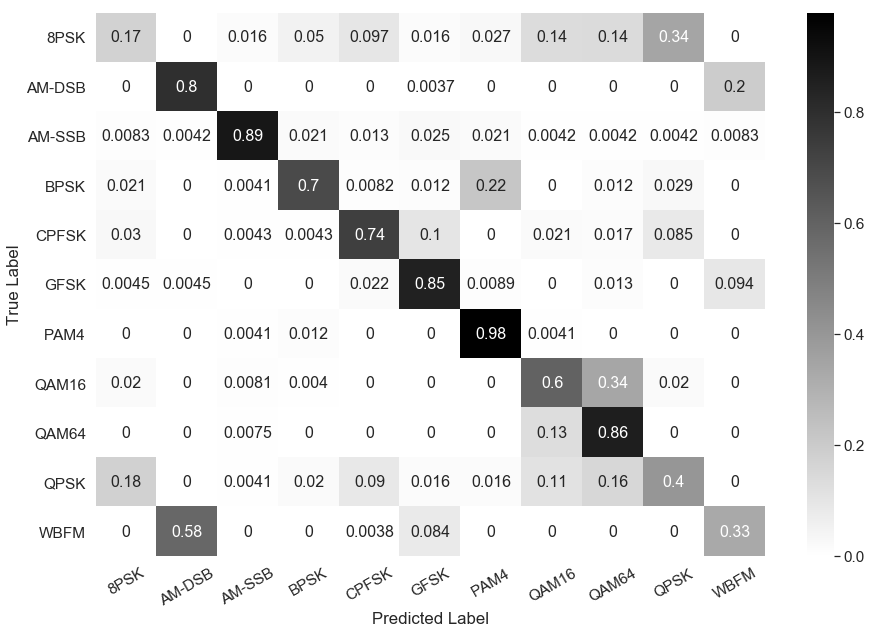}
	\caption{Confusion matrix for the proposed model (top) and LSTM (bottom) on RadioML2016.10A dataset at -4dB SNR.}
	\label{fig:confusion_3}
\end{figure}

The two-layer LSTM denoising auto-encoder is trained on SNRs ranging from -20dB to 18dB. The top-1 classification 
accuracy of CNN \cite{Shea2016}, CLDNN \cite{West2017}, LSTM \cite{Rajendran2018} and our  proposed model is 
shown in Figure~\ref{fig:SNR} and Table~\ref{tab:10a_result}. The classification accuracy computed over all SNRs 
achieved by CNN, CLDNN, LSTM and our proposed model is 51.29\%, 56.78\%, 60.49\% and 61.72\%, respectively. 
It is worth pointing out that training with noise helps increase the classification accuracy (computed across all SNRs) 
by 1.1\% as compared to training with the original signal. The auto-encoder enables extraction of stable low-dimensional 
features with a significantly reduced dimension of hidden LSTM states and thus contributes to the improvement in
classification accuracy and the reduction of computational complexity.
The average classification 
accuracy for SNRs ranging from 0dB to 18dB achieved by CNN, CLDNN, LSTM and our proposed model is 73.9\%, 
83.31\%, 90.26\% and 91.55\%, respectively. The proposed model outperforms selected models for almost all SNRs. 
The considered softwares were executed with their default settings, i.e., we use the same hyperparameters as the
authors of existing methods did when running their methods on RadioML10.A dataset. The results in 
Figure~\ref{fig:SNR} are averaged over 10 experiments. Model parameters were initialized at the beginning of each 
experiment. Note that the benchmarking results for the pre-existing methods that we obtained closely match those 
reported in \cite{Rajendran2018}. It is worth pointing out that 
our proposed model significantly outperforms CLDNN and CNN in high SNR regimes, while marginally outperforming
LSTM in terms of top-1 classification accuracy. Figure~\ref{fig:confusion_1}-\ref{fig:confusion_3} illustrate the confusion matrices in the experiment that achieved the highest 
overall top-1 classification accuracy for the proposed model and LSTM for SNRs 18dB, 0dB and -4dB. For the SNR of 18dB,
the diagonal is much more sharp even though there are some confusions in separating AM-DSB from WBFM signals, which 
are mainly due to the silence periods of audio \cite{Rajendran2018}. Similar to Figure~\ref{fig:confusion_2}, there are some 
difficulties in separating AM-DSB and WBFM at the SNR of 0dB. Besides, there is some level of confusion between QAM16 
and QAM64 since QAM16 is a subset of QAM64. It is also worth mentioning that the proposed model performs better on 
AM-SSB signals and on distinguishing QAM64 from QAM16 signals at high SNRs. As shown in Figure~\ref{fig:confusion_3}, 
it becomes much more difficult to distinguish the signals at low SNRs. 

\begin{table}[!ht]
\caption{A comparison of the number of trainable parameters, FLOPs and memory cost on RadioML2016.10A data.}
\centering
{\begin{tabular}{>{\centering\arraybackslash}p{0.1\columnwidth} >{\centering\arraybackslash}p{0.22\columnwidth} >{\centering\arraybackslash}p{0.22\columnwidth} >{\centering\arraybackslash}p{0.22\columnwidth}}
\toprule
Model  & \# Parameters&\# FLOPs & Memory \\
\midrule
DAE & 14637 & 45040  &224KB\\
LSTM         & 200075  & 660239  &2.32MB\\
CLDNN      &  248817 & 588974  & 2.91MB\\
CNN           & 5456219  & 80548043  &61.4MB\\
\bottomrule
\end{tabular}}
\label{tab:parameters_1}
\end{table}

\begin{table}[!ht]
\caption{Number of classifications per second on different platforms on RadioML2016.10A data.}
\centering
{\begin{tabular}{>{\centering\arraybackslash}p{0.25\columnwidth} >{\centering\arraybackslash}p{0.07\columnwidth} >{\centering\arraybackslash}p{0.09\columnwidth} >{\centering\arraybackslash}p{0.09\columnwidth} >{\centering\arraybackslash}p{0.09\columnwidth} >{\centering\arraybackslash}p{0.13\columnwidth}}
\toprule
Platform &  & DAE & LSTM & CLDNN & CNN \\
\midrule
GTX 1080Ti & Mean & 8456 & 7325 & 10257 & 19162 \\
& Std & 153.97 & 138.72  & 102.28 &  346.64 \\
Intel i7-8700K & Mean & 1255 & 869  & 871 & 1578 \\
& Std & 7.72  &  21.62  &  15.70 & 14.36\\
Raspberry Pi 4& Mean  & 241  & 42 & 45  & 127  \\
& Std & 14.61 & 0.51  & 1.36 &  1.01 \\
Raspberry Pi 3 & Mean  & 119 & 19 & 20 & 45  \\
& Std & 1.45 & 1.97  & 0.23 & 2.39 \\
\bottomrule
\end{tabular}}
\label{tab:speed_1}
\end{table}

Next, we compare considered models in terms of the number of trainable parameters, the number of floating point 
operations (FLOPs), the memory cost, and the number of classifications per second on different computational platforms. 
As shown in Table~\ref{tab:parameters_1}, the proposed model has the smallest number of trainable parameters and 
requires the fewest FLOPs and the smallest memory space. Table~\ref{tab:speed_1} shows that in terms of the number 
of classifications per second on Raspberry Pi 4, the proposed model is on average approximately $5.6\times$, $5.4\times$ 
and $1.9\times$ faster than LSTM, CLDNN and CNN, respectively. On Raspberry Pi 3, the proposed model is on average 
approximately $6.3\times$, $6\times$ and $2.6\times$ faster than LSTM, CLDNN and CNN, respectively. The mean and 
standard deviation of the number of classifications per second are averaged over 10 experiments. Note that the complexity 
of existing methods cannot be reduced without causing severe deterioration of classification accuracy \cite{Rajendran2018}.

\subsection{Performance Comparison on RadioML2018.01A}
We next evaluate performance of the proposed model on the modulation classification task using the realistic over-the-air 
RadioML2018.01A data with specific radio channel effect settings including carrier frequency offset, symbol rate offset, 
delay spread and thermal noise \cite{Oshea2018}. Signals over the so-called Normal Classes that are commonly seen in 
impaired environments, including OOK, 4ASK, BPSK, QPSK, 8PSK, 16QAM, AM-SSB-SC, AM-DSB-SC, FM, GMSK and 
OQPSK, are utilized. The data contains 11 modulations and the SNR range is from -20dB to 30dB with 2dB step size. For 
each SNR and modulation scheme there are 4096 samples, leading to about 1.17M samples in total. The sample length is 
1024 and each sample is composed of IQ components. 50\%, 25\% and 25\% of the entire dataset are used for training, 
validation and testing, respectively. 

Table~\ref{tab:01a_result} shows the mean and standard deviation of top-1 classification accuracy over the range of SNRs 
for a number of models computed over 10 experiments. The classification accuracy computed over all SNRs achieved by 
VGG, RN and our proposed model is 64.03\%, 66.00\% and 67.30\%, respectively. 
Note that training with noise helps increase the classification accuracy (computed across all SNRs) by 0.9\% as compared 
to training with the original signal. As before, the auto-encoder enables extraction of stable low-dimensional features with 
a significantly reduced dimension of hidden LSTM states, hence contributing to the improvement in classification accuracy 
and the reduction of computational complexity. The classification accuracy over the range of SNR from $0$dB to 
$30$dB achieved by VGG, RN and our proposed model is 92.16\%, 94.89\% and 96.56\%, respectively. The proposed model 
outperforms state-of-the-art models for almost all SNRs.

\begin{figure}[H]
	\centering
	\includegraphics[width= 0.9\columnwidth]{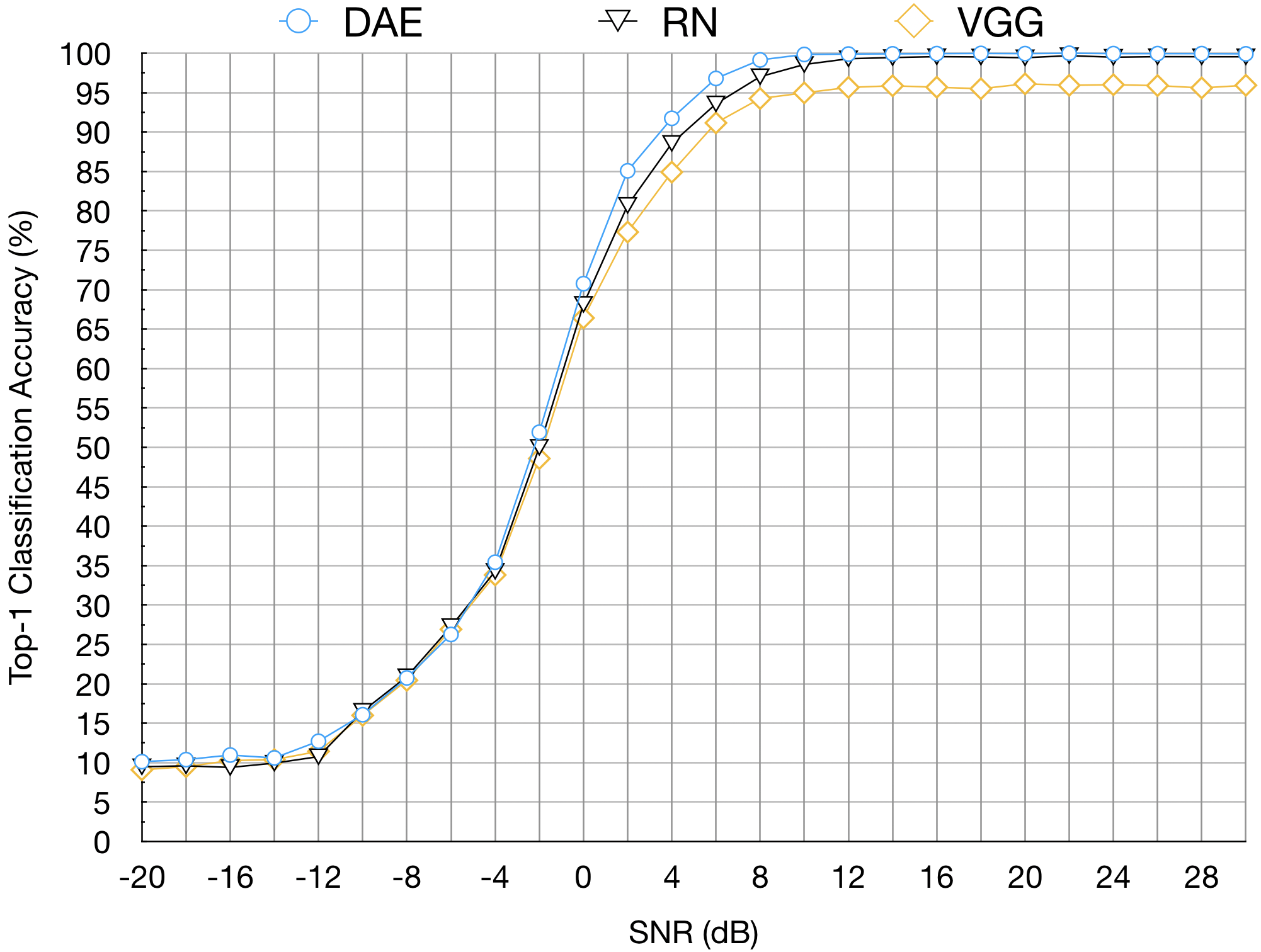}
	\caption{Top-1 classification accuracy comparison of the proposed model (DAE) with selected models on RadioML2018.01A dataset.}\label{fig:SNR_2}
\end{figure}

The top-1 classification accuracy of the VGG and residual networks (RN) used in \cite{Oshea2018} and the proposed model across different 
SNRs are shown in Figure~\ref{fig:SNR_2}. Results in Figure~\ref{fig:SNR_2} are averaged over 10 experiments.

\begin{figure}[H]
	\centering
	\includegraphics[width= 0.6\columnwidth]{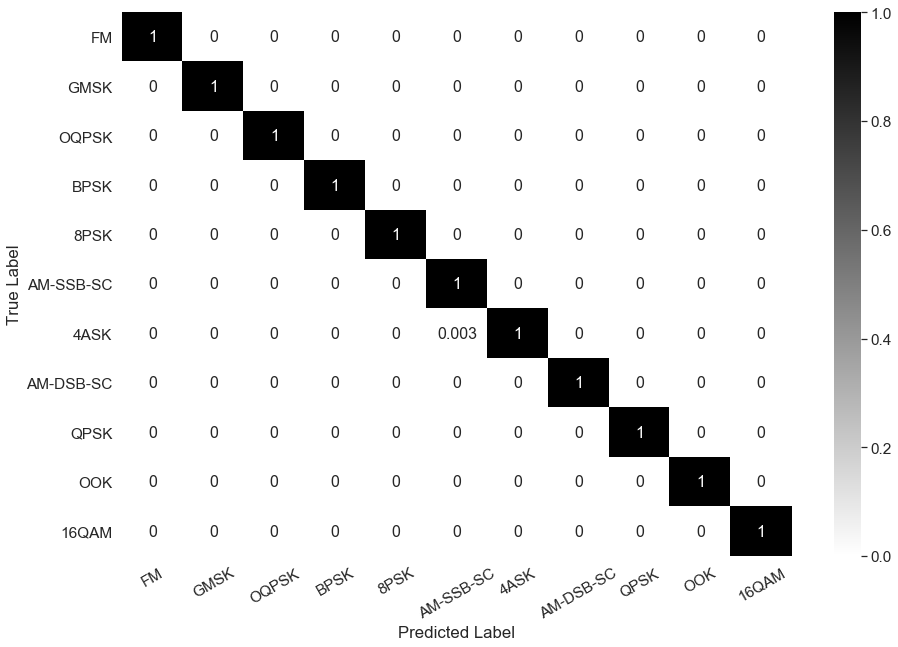}
	\includegraphics[width= 0.6\columnwidth]{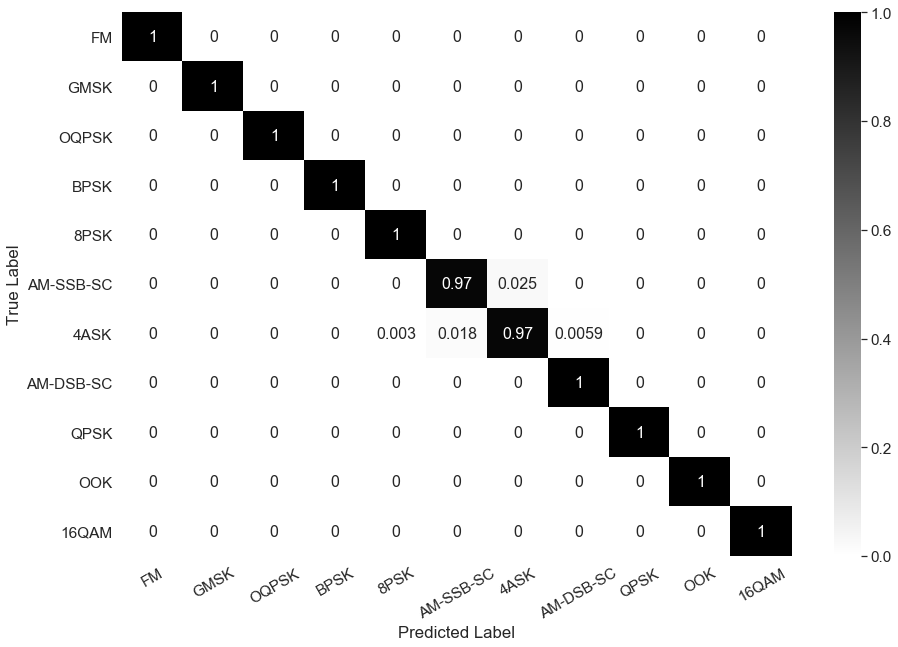}
	\includegraphics[width= 0.6\columnwidth]{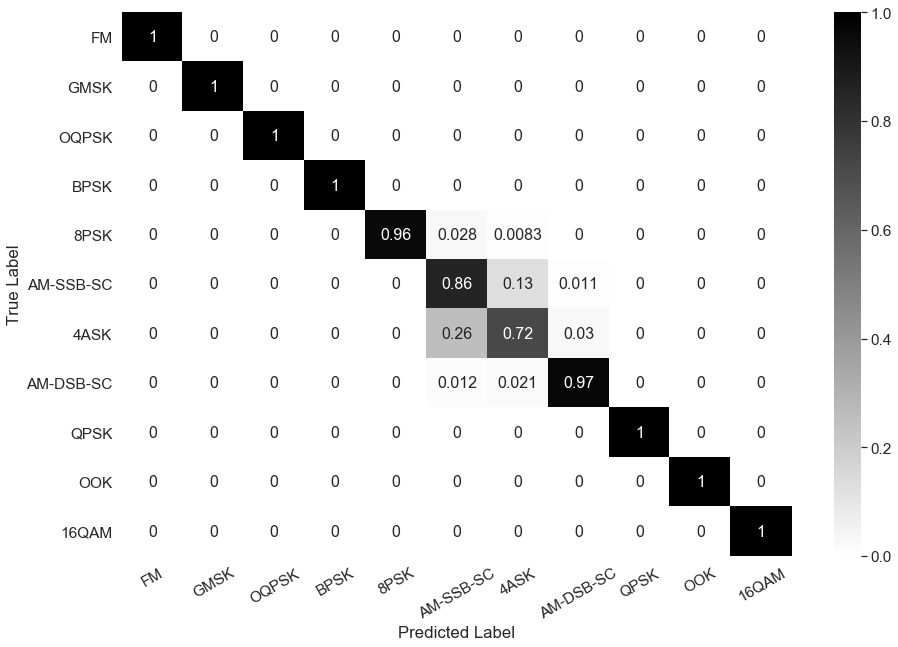}
	\caption{Confusion matrix for the proposed model (top), RN (middle) and VGG (bottom) on RadioML2018.01A dataset at 18dB SNR.}
	\label{fig:confusion_4}
\end{figure}

\begin{figure}[H]
	\centering
	\includegraphics[width= 0.6\columnwidth]{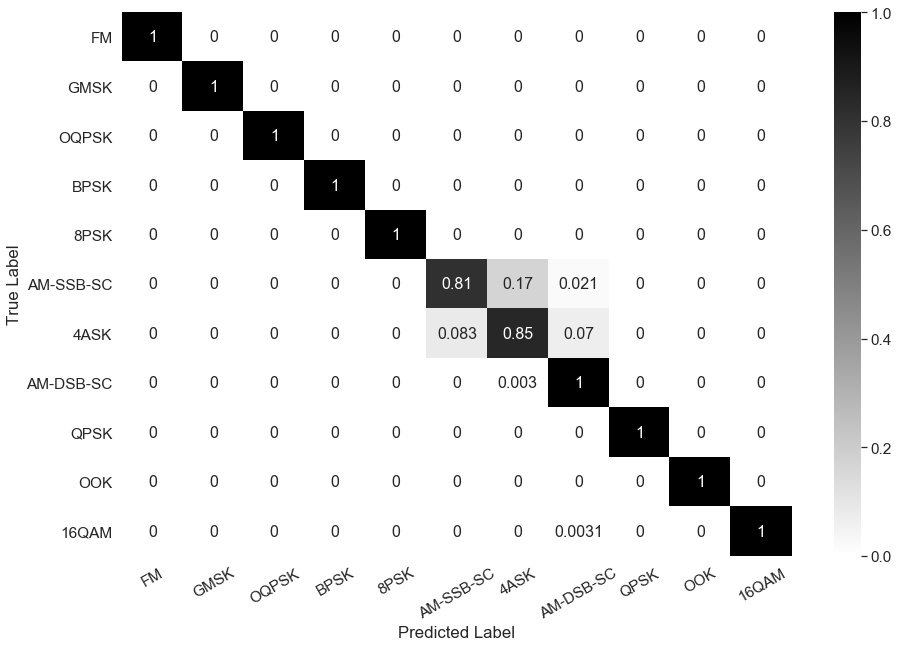}
	\includegraphics[width= 0.6\columnwidth]{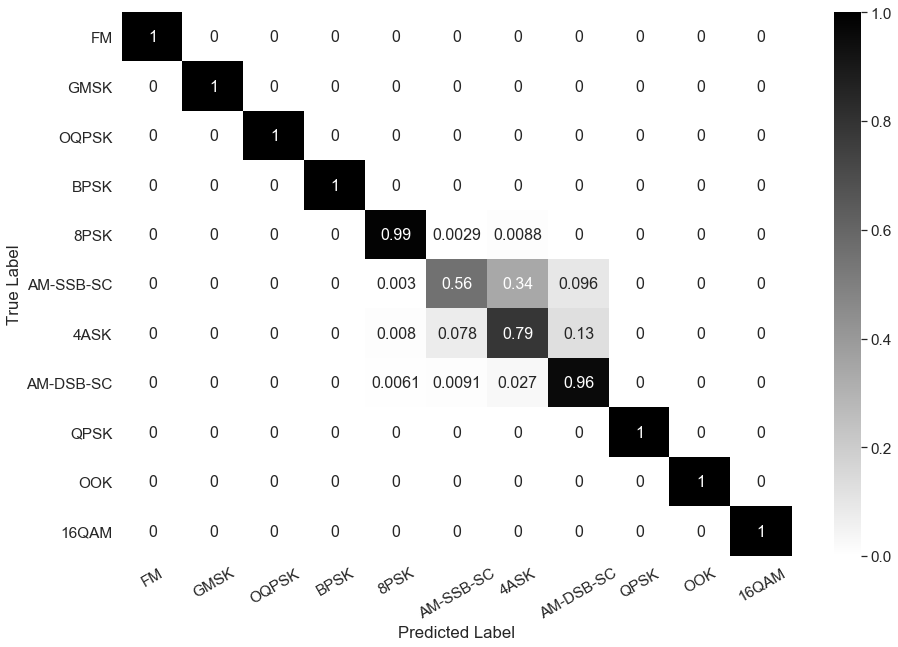}
	\includegraphics[width= 0.6\columnwidth]{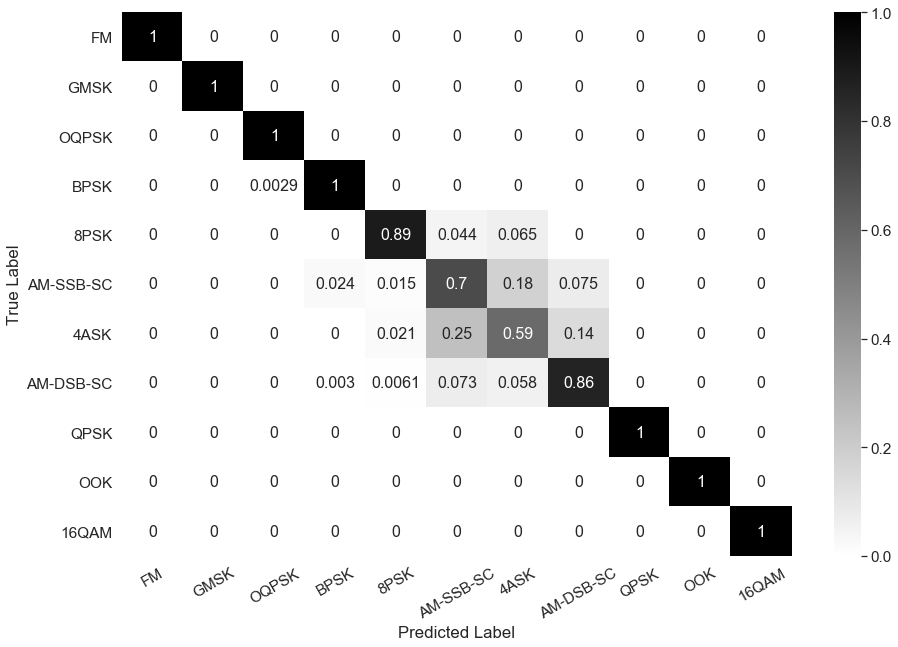}
	\caption{Confusion matrix for the proposed model (top), RN (middle) and VGG (bottom) on RadioML2018.01A dataset at 6dB SNR.}
	\label{fig:confusion_5}
\end{figure}

\begin{figure}[H]
	\centering
	\includegraphics[width= 0.6\columnwidth]{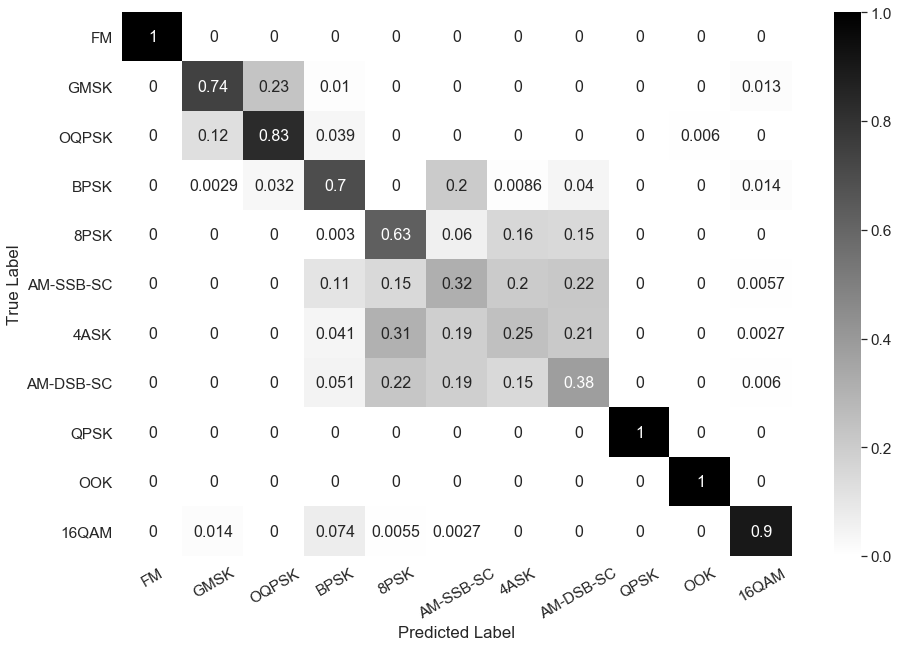}
	\includegraphics[width= 0.6\columnwidth]{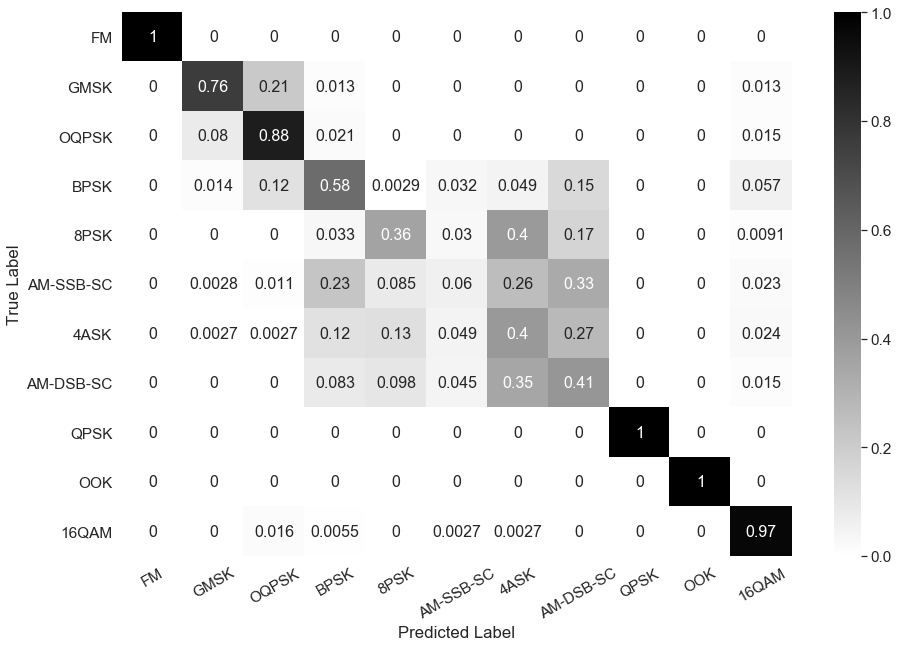}
	\includegraphics[width= 0.6\columnwidth]{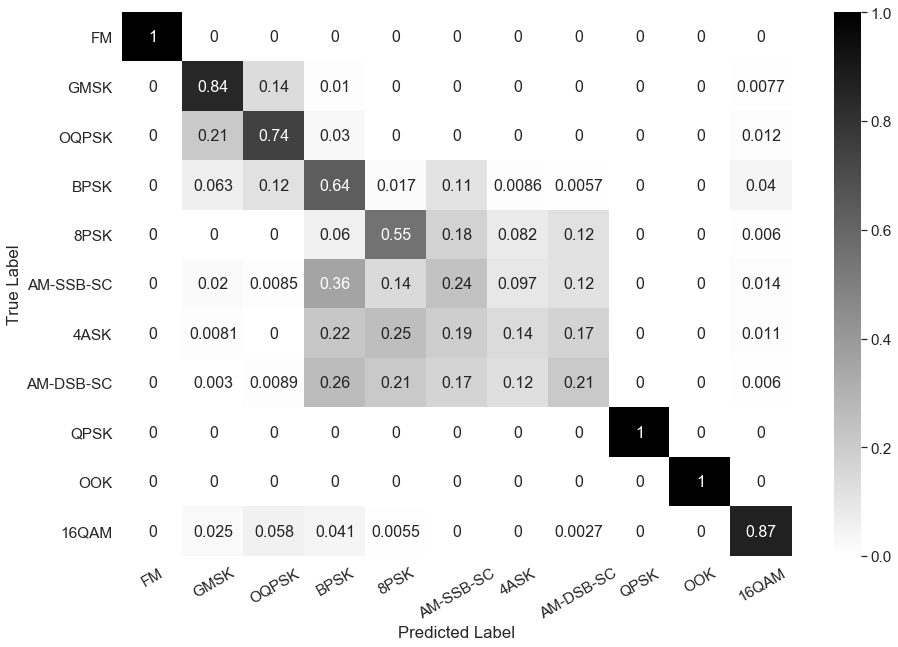}
	\caption{Confusion matrix for the proposed model (top), RN (middle) and VGG (bottom) on RadioML2018.01A dataset at 0dB SNR.}
	\label{fig:confusion_6}
\end{figure}

Figure~\ref{fig:confusion_4}-\ref{fig:confusion_6} illustrate the confusion matrices for the experiment with the highest overall top-1 classification 
accuracy for the proposed model and LSTM at the SNRs 18dB, 6dB and 0dB. For the SNR of 18dB, the diagonal is very sharp for the proposed 
model while there are some confusions between AM-SSB-SC and 4ASK signals for RN and VGG. At the SNR of 0dB, it becomes more difficult 
for the proposed model to separate AM-SSB-SC and 4ASK. As shown in Figure~\ref{fig:confusion_6}, it becomes much more difficult to distinguish the signals at low SNRs and all considered models start making mistakes differentiating between GMSK, OQPSK and BPSK signals.


\begin{table*}[!ht]
\caption{Top-1 classification accuracy comparison of the proposed model (DAE) vs. existing models 
on RadioML2018.01A dataset (mean and standard deviation). The highest average top-1 classification 
accuracy for each SNR is marked in bold.}\label{tab:01a_result}
\centering
\resizebox{0.95\columnwidth}{!}{\begin{tabular}{>{\centering\arraybackslash}p{0.1\columnwidth} >{\centering\arraybackslash}p{0.1\columnwidth} >{\centering\arraybackslash}p{0.09\columnwidth}>{\centering\arraybackslash}p{0.09\columnwidth} >{\centering\arraybackslash}p{0.09\columnwidth} >{\centering\arraybackslash}p{0.09\columnwidth} >{\centering\arraybackslash}p{0.09\columnwidth} >{\centering\arraybackslash}p{0.09\columnwidth} >{\centering\arraybackslash}p{0.09\columnwidth}}
\toprule
Model  &  & -20dB & -18dB & -16dB & -14dB & -12dB & -10dB & -8dB \\
\midrule
DAE & Mean & \textbf{9.86} & \textbf{9.73} &  \textbf{10.03} &  \textbf{11.05} &  \textbf{12.84} &\textbf{17.33} &  \textbf{20.43} \\
& Std & 0.32 & 0.32 & 0.40 & 0.48 &  0.42&0.46  & 0.43 \\
RN & Mean & 9.28 & 9.42 &  9.69 &  10.49 &  11.81 &16.21 &  19.87\\
& Std & 0.50 & 0.60 & 0.46 &  0.43& 0.89  & 0.77 & 0.87   \\
VGG & Mean & 9.82 & 9.17 &  9.69 &  10.59 &  11.50 &15.41 &  19.87\\
& Std & 0.47 & 0.40 & 0.38 & 0.41 &  0.48&0.67  & 0.39 \\
\bottomrule
\end{tabular}}

\vspace{5pt}

\resizebox{0.95\columnwidth}{!}{\begin{tabular}{>{\centering\arraybackslash}p{0.1\columnwidth} >{\centering\arraybackslash}p{0.1\columnwidth} >{\centering\arraybackslash}p{0.09\columnwidth}>{\centering\arraybackslash}p{0.09\columnwidth} >{\centering\arraybackslash}p{0.09\columnwidth} >{\centering\arraybackslash}p{0.09\columnwidth} >{\centering\arraybackslash}p{0.09\columnwidth} >{\centering\arraybackslash}p{0.09\columnwidth} >{\centering\arraybackslash}p{0.09\columnwidth}}
\toprule
Model  &  & -6dB & -4dB & -2dB & 0dB & 2dB & 4dB & 6dB\\
\midrule
DAE & Mean & 26.71&  \textbf{35.84}  & \textbf{52.41} & \textbf{72.04} &  \textbf{85.31} & \textbf{ 92.10} &  \textbf{97.19}\\
& Std &0.32  & 0.43 & 0.51 & 0.26 & 0.32 & 0.38 &  0.16\\
RN & Mean &  26.54&  34.33 & 50.34 & 68.65 &  80.90 &  87.88 &  93.54 \\
& Std  &  1.35 & 0.95 & 0.98 & 1.21 & 0.72 & 0.69 &  0.50\\
VGG & Mean &  \textbf{27.69}&  34.57 & 47.91 & 65.43 &  78.04 &  85.29 &  91.25\\
& Std  &0.46  & 0.66 & 0.82 & 0.65 & 0.62 & 0.56 &  0.20\\
\bottomrule
\end{tabular}}

\vspace{5pt}

\resizebox{0.95\columnwidth}{!}{\begin{tabular}{>{\centering\arraybackslash}p{0.1\columnwidth} >{\centering\arraybackslash}p{0.1\columnwidth} >{\centering\arraybackslash}p{0.09\columnwidth}>{\centering\arraybackslash}p{0.09\columnwidth} >{\centering\arraybackslash}p{0.09\columnwidth} >{\centering\arraybackslash}p{0.09\columnwidth} >{\centering\arraybackslash}p{0.09\columnwidth} >{\centering\arraybackslash}p{0.09\columnwidth} >{\centering\arraybackslash}p{0.09\columnwidth}}
\toprule
Model  &  & 8dB & 10dB & 12dB & 14dB & 16dB & 18dB & 20dB\\
\midrule
DAE & Mean &\textbf{99.33} &  \textbf{99.77}&  \textbf{99.91}&  \textbf{99.87} & \textbf{99.95} & \textbf{99.95} &  \textbf{99.91}\\
& Std &0.09  & 0.05 &0.05  & 0.05 & 0.05 & 0.08 & 0.05\\
RN & Mean &  97.26 &  98.29 &  98.81 &  99.12 & 99.26 & 99.24 &  99.28 \\
& Std  & 0.28  & 0.41 & 0.50 & 0.29 &  0.32 & 0.37 & 0.30\\
VGG & Mean &94.07 &  94.75&  95.80&  95.89 & 96.10 & 95.70 &  96.22\\
& Std  &0.23  & 0.31 &0.32  & 0.38 & 0.34 & 0.29 & 0.24\\
\bottomrule
\end{tabular}}

\vspace{5pt}

\resizebox{0.95\columnwidth}{!}{\begin{tabular}{>{\centering\arraybackslash}p{0.1\columnwidth} >{\centering\arraybackslash}p{0.1\columnwidth} >{\centering\arraybackslash}p{0.09\columnwidth}>{\centering\arraybackslash}p{0.09\columnwidth} >{\centering\arraybackslash}p{0.09\columnwidth} >{\centering\arraybackslash}p{0.09\columnwidth} >{\centering\arraybackslash}p{0.09\columnwidth} >{\centering\arraybackslash}p{0.09\columnwidth} >{\centering\arraybackslash}p{0.09\columnwidth}}
\toprule
Model  &  & 22dB& 24dB & 26dB & 28dB & 30dB & Overall \\
\midrule
DAE & Mean & \textbf{99.94} & \textbf{99.97}& \textbf{99.95} &  \textbf{99.91}&  \textbf{99.93} &  \textbf{67.30}  & \\
& Std & 0.01 &  0.03 &  0.06 &  0.04 &  0.02 &  0.17 & \\
RN & Mean  &  99.21 &  99.17 &  99.22 &  99.15 &  99.28 &  66.00 & \\
& Std  & 0.38 &  0.39 &  0.35 &  0.36&  0.40 &  0.22\\
VGG &  Mean &96.09 &  96.16 &  95.97 &  96.15&  96.10 &  64.03 &  \\
& Std & 0.42 &  0.39&  0.24 &  0.36&  0.31 &  0.15 &  \\
\bottomrule
\end{tabular}}
\end{table*}

In addition to the top-1 classification accuracy, we also compare considered models in terms of the number 
of trainable parameters, the number of floating point operations, the memory cost and the number 
of classifications per second on different computational platforms. As shown in Table~\ref{tab:parameters_3}, 
the proposed model has the fewest trainable parameters and requires the smallest number of FLOPs and 
memory space. It is noticeable in Table~\ref{tab:speed_3} that the proposed model is on average approximately 
$2.4\times$ and $1.6\times$ faster than RN and VGG on Raspberry Pi 4 in terms of the number of classifications 
per second, respectively. On Raspberry Pi 3, the proposed model is on average approximately $2.4\times$ and 
$1.3\times$ faster than RN and VGG, respectively. The mean and standard deviation of the number of 
classifications per second are calculated over 10 experiments on 1024 signals.

\begin{table}[!ht]
\caption{Number of trainable parameters, number of FLOPs and memory cost of the considered models on RadioML2018.01A data.}
\centering
{\begin{tabular}{>{\centering\arraybackslash}p{0.1\columnwidth} >{\centering\arraybackslash}p{0.22\columnwidth} >{\centering\arraybackslash}p{0.22\columnwidth} >{\centering\arraybackslash}p{0.22\columnwidth}}
\toprule
Model  & \# Parameters&\# FLOPs & Memory \\
\midrule
DAE & 14989 & 288925  & 242KB\\
RN         &  257009 & 8651090  & 3.41MB\\
VGG      &  236344 & 8343647  & 3.42MB\\
\bottomrule
\end{tabular}}
\label{tab:parameters_3}
\end{table}

\begin{table}[!ht]
\caption{Number of classifications per second on different platforms on RadioML2018.01A data.}
\centering
{\begin{tabular}{>{\centering\arraybackslash}p{0.25\columnwidth} >{\centering\arraybackslash}p{0.07\columnwidth} >{\centering\arraybackslash}p{0.11\columnwidth} >{\centering\arraybackslash}p{0.11\columnwidth} >{\centering\arraybackslash}p{0.11\columnwidth} }
\toprule
Platform &  & DAE & RN & VGG  \\
\midrule
GTX 1080Ti & Mean & 1327.45  & 902.71 & 1326.73  \\
& Std & 18.49 & 22.22  & 16.64  \\
Intel i7-8700K & Mean & 119.25  &  85.73 &  155.77 \\
& Std & 1.66  &  2.36  &  1.63 \\
Raspberry Pi 4& Mean  & 24.43  & 10.32 &  15.43  \\
& Std & 0.26  & 0.36  & 0.61  \\
Raspberry Pi 3 & Mean  & 12.12 & 5.28 & 9.06 \\
& Std & 0.17 & 0.40  & 0.43 \\
\bottomrule
\end{tabular}}
\label{tab:speed_3}
\end{table}

\subsection{Performance Comparison on Electrosense Data}
We further evaluate performance of the proposed model on real-time over-the-air PSD data from Electrosense. 
The goal of Electrosense initiative is to enable more efficient, safe and reliable monitoring of the electromagnetic 
space by improving accessibility of spectrum data to general public \cite{SRajendran2018}. The aggregated 
spectrum measurements collected from sensors all over the world could be retrieved from the Electrosense 
API\footnote{\url{https://electrosense.org/open-api-spec.html}}. Six commercially deployed technologies 
(WFM, TETRA, DVB, RADAR, LTE and GSM) are collected from indoor sensors with omni-directional antennas 
by setting frequency resolution to 100kHz and time resolution to 60s \cite{Rajendran2018}. 10k samples of length 
2000 are retrieved for each technology and are padded with 0s accordingly for the consistency of the sample 
lengths. 50\%, 25\% and 25\% of the entire dataset are used for training, validation and testing, respectively.
\begin{figure}[H]
	\centering
	\includegraphics[width= 0.65\columnwidth]{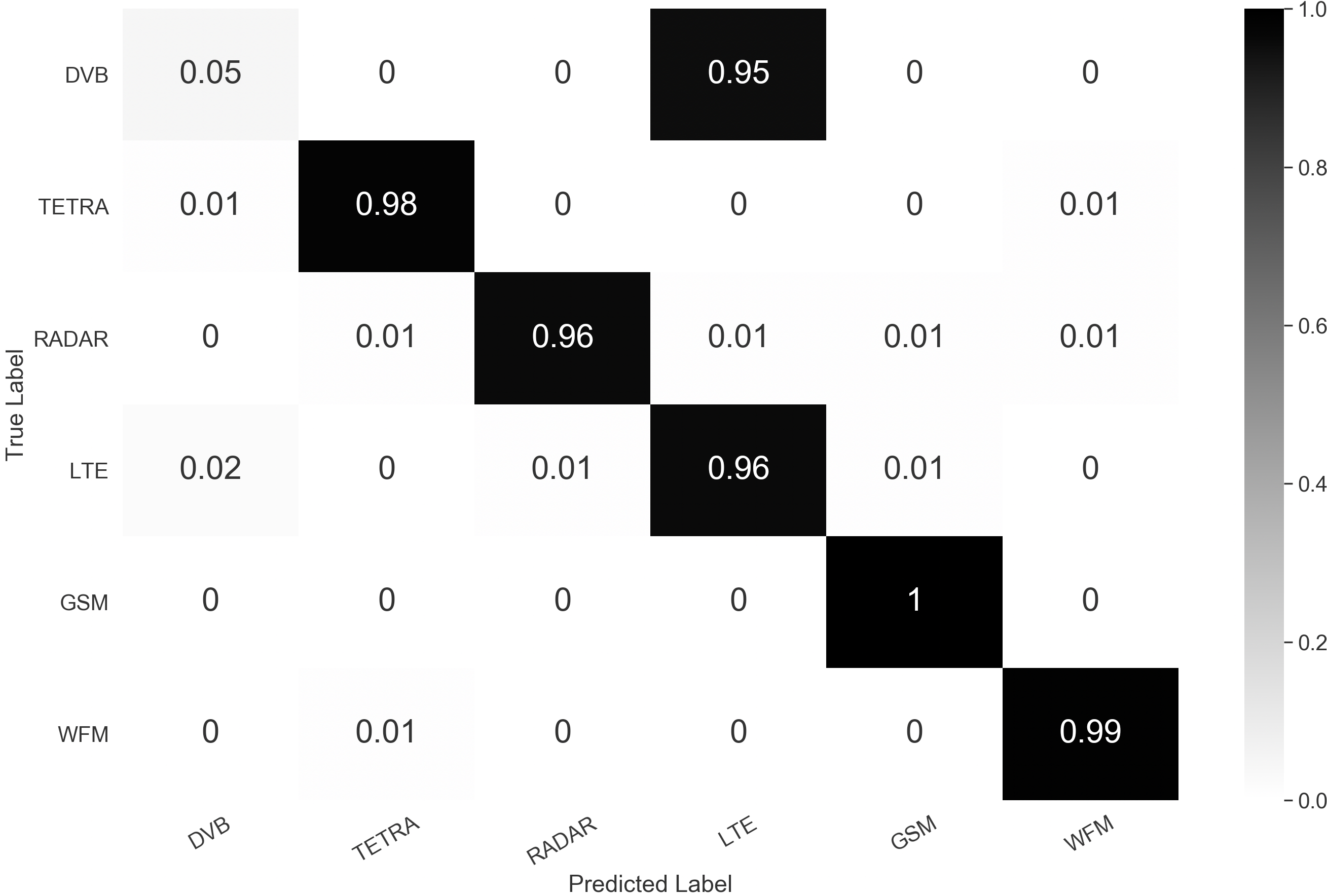}
	\includegraphics[width= 0.65\columnwidth]{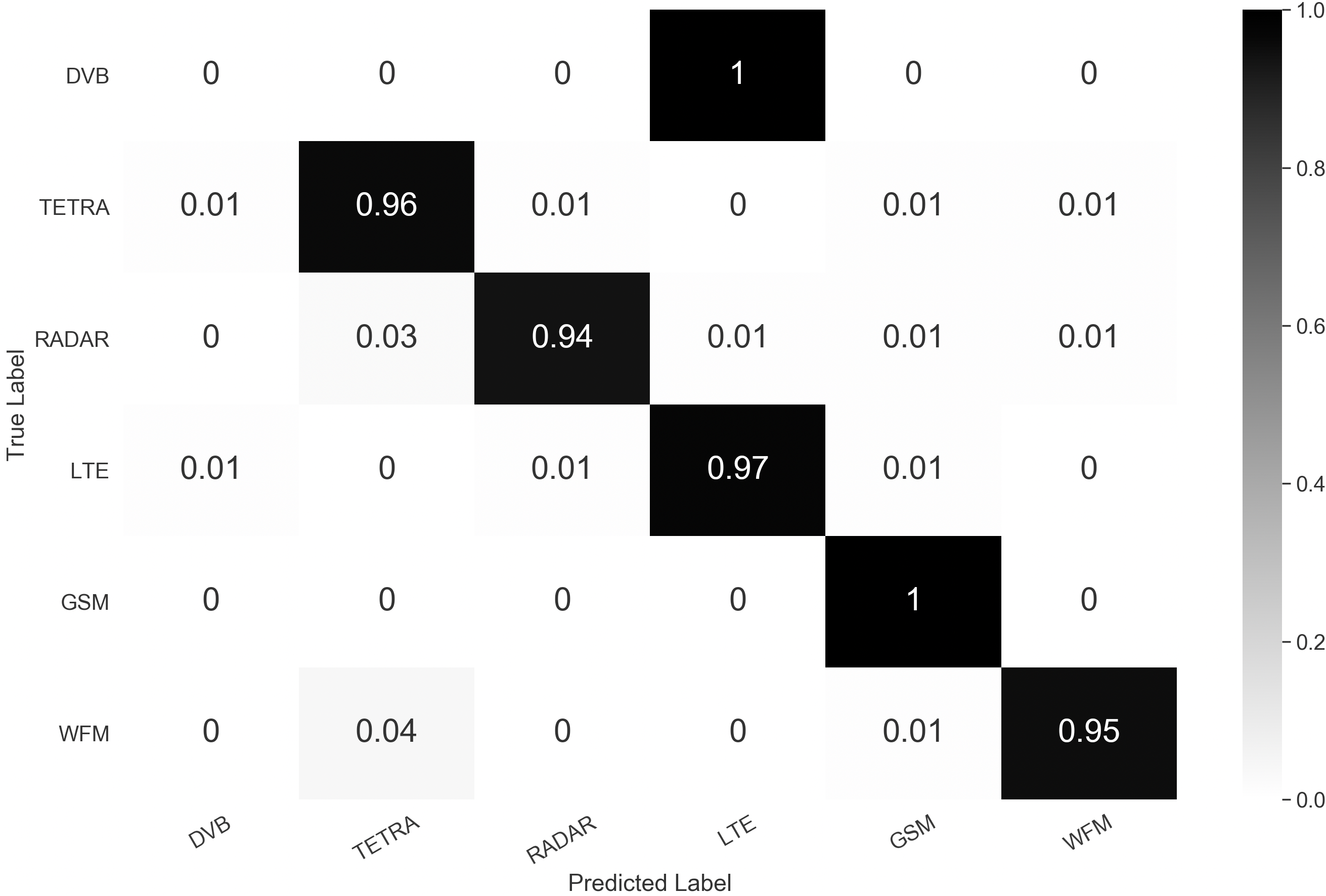}
	\caption{Confusion matrices of the proposed model (top) and LSTM (bottom) for technology classification 
	on Electrosense data.}
	\label{fig:confusion_7}
\end{figure}

Figure~\ref{fig:confusion_7} shows the confusion matrices of our proposed model and LSTM for technology
 classification on Electrosense data. The proposed model performs slightly better than LSTM. It is noticeable 
 that distinguishing DVB from LTE based on PSD is difficult since the power spectra of DVB and LTE is highly 
 similar and both of them are based on OFDM \cite{Rajendran2018}.

Next, we compare the considered models in terms of the number of trainable parameters, the number of FLOPs, 
the memory cost and the number of classifications per second on different platforms. As shown in 
Table~\ref{tab:parameter_2}, the proposed model has significantly fewer trainable parameters and requires much 
fewer FLOPs and memory space. Table~\ref{tab:speed_2} shows that the proposed model is on average 
approximately $4.3\times$ faster than LSTM on Raspberry Pi 4 in terms of the number of classifications per second. 
On Raspberry Pi 3, the proposed model is on average approximately $6\times$ faster than LSTM. The mean and 
standard deviation of the number of classifications per second are calculated over 10 experiments on 1024 signals.

\begin{table}[!ht]
\caption{Number of trainable parameters, number of FLOPs and memory cost of the considered models on Electrosense data.}
\centering
{\begin{tabular}{>{\centering\arraybackslash}p{0.1\columnwidth} >{\centering\arraybackslash}p{0.22\columnwidth} >{\centering\arraybackslash}p{0.22\columnwidth} >{\centering\arraybackslash}p{0.22\columnwidth}}
\toprule
Model  & \# Parameters&\# FLOPs & Memory \\
\midrule
DAE & 14572 & 279455  & 237KB\\
LSTM   & 199563  & 7696283  &2.31MB\\
\bottomrule
\end{tabular}}
\label{tab:parameter_2}
\end{table}

\begin{table}[!ht]
\caption{Number of classifications per second on different platforms for Electrosense data.}
\centering
{\begin{tabular}{>{\centering\arraybackslash}p{0.25\columnwidth} >{\centering\arraybackslash}p{0.07\columnwidth} >{\centering\arraybackslash}p{0.11\columnwidth} >{\centering\arraybackslash}p{0.11\columnwidth}}
\toprule
Platform &  & DAE & LSTM \\
\midrule
GTX 1080Ti & Mean & 685.08  & 613.30 \\
& Std & 13.88 & 10.85 \\
Intel i7-8700K & Mean & 77.72 &  64.08 \\
& Std & 1.69 & 0.76 \\
Raspberry Pi 4& Mean  & 12.79  &  2.92 \\
& Std & 0.04 & 0.02 \\
Raspberry Pi 3 & Mean  & 6.45 &  1.27\\
& Std &  0.06 & 0.01 \\
\bottomrule
\end{tabular}}
\label{tab:speed_2}
\end{table}

\section{Conclusions}
In this paper, we introduce a denoising auto-encoder to the problem of inferring the 
modulation and technology type of a received radio signal. 
In particular, an LSTM 
auto-encoder is trained to learn stable and robust features from the noise corrupted 
received signals, reconstruct the original received signals and infer the modulation or technology type, simultaneously. Empirical studies show that the proposed framework generally outperforms top-1 classification accuracy of the competing methods while
requiring significantly smaller computation resources. In particular, the proposed 
framework employs a compact architecture that it can be implemented on affordable computational devices, enabling real-time classification of the received signals at
required levels of accuracy.


%


%
%
%

\ifCLASSOPTIONcaptionsoff
  \newpage
\fi



\bibliographystyle{IEEEtran}
\bibliography{IEEEabrv,IEEEtran}

\begin{thebibliography}{10}
\providecommand{\url}[1]{#1}
\csname url@samestyle\endcsname
\providecommand{\newblock}{\relax}
\providecommand{\bibinfo}[2]{#2}
\providecommand{\BIBentrySTDinterwordspacing}{\spaceskip=0pt\relax}
\providecommand{\BIBentryALTinterwordstretchfactor}{4}
\providecommand{\BIBentryALTinterwordspacing}{\spaceskip=\fontdimen2\font plus
\BIBentryALTinterwordstretchfactor\fontdimen3\font minus
  \fontdimen4\font\relax}
\providecommand{\BIBforeignlanguage}[2]{{%
\expandafter\ifx\csname l@#1\endcsname\relax
\typeout{** WARNING: IEEEtran.bst: No hyphenation pattern has been}%
\typeout{** loaded for the language `#1'. Using the pattern for}%
\typeout{** the default language instead.}%
\else
\language=\csname l@#1\endcsname
\fi
#2}}
\providecommand{\BIBdecl}{\relax}
\BIBdecl

\bibitem{Dobre2007}
O.~A. {Dobre}, A.~{Abdi}, Y.~{Bar-Ness}, and W.~{Su}, ``Survey of automatic
  modulation classification techniques: classical approaches and new trends,''
  \emph{IET Communications}, vol.~1, no.~2, pp. 137--156, 2007.

\bibitem{Gardner1988}
W.~A. {Gardner}, ``Signal interception: a unifying theoretical framework for
  feature detection,'' \emph{IEEE Transactions on Communications}, vol.~36,
  no.~8, pp. 897--906, 1988.

\bibitem{Gardner1992}
W.~A. {Gardner} and C.~M. {Spooner}, ``Signal interception: performance
  advantages of cyclic-feature detectors,'' \emph{IEEE Transactions on
  Communications}, vol.~40, no.~1, pp. 149--159, 1992.

\bibitem{Gardner19871}
W.~{Gardner}, ``Spectral correlation of modulated signals: Part i - analog
  modulation,'' \emph{IEEE Transactions on Communications}, vol.~35, no.~6, pp.
  584--594, 1987.

\bibitem{Gardner19872}
W.~{Gardner}, W.~{Brown}, and {Chih-Kang Chen}, ``Spectral correlation of
  modulated signals: Part ii - digital modulation,'' \emph{IEEE Transactions on
  Communications}, vol.~35, no.~6, pp. 595--601, 1987.

\bibitem{Yu2006}
Z.~Yu, ``Automatic modulation classification of communication signals,'' Ph.D.
  dissertation, Department of Electrical and Computer Engineering, New Jersey
  Institute of Technology, 2006.

\bibitem{Fehske2005}
A.~{Fehske}, J.~{Gaeddert}, and J.~H. {Reed}, ``A new approach to signal
  classification using spectral correlation and neural networks,'' \emph{First
  IEEE International Symposium on New Frontiers in Dynamic Spectrum Access
  Networks, 2005. DySPAN 2005.}, no. 144-150, 2005.

\bibitem{Shea2016}
T.~J. O'Shea, J.~Corgan, and T.~C. Clancy, ``Convolutional radio modulation
  recognition networks,'' \emph{Engineering Applications of Neural Networks},
  pp. 213--226, 2016.

\bibitem{West2017}
N.~E. {West} and T.~{O'Shea}, ``Deep architectures for modulation
  recognition,'' \emph{2017 IEEE International Symposium on Dynamic Spectrum
  Access Networks (DySPAN)}, pp. 1--6, 2017.

\bibitem{Hochreiter1997}
S.~Hochreiter and J.~Schmidhuber, ``Long short-term memory,'' \emph{Neural
  computation}, vol.~9, pp. 1735--80, 12 1997.

\bibitem{Gers1999}
F.~A. {Gers}, J.~{Schmidhuber}, and F.~{Cummins}, ``Learning to forget:
  continual prediction with lstm,'' \emph{1999 Ninth International Conference
  on Artificial Neural Networks ICANN 99. (Conf. Publ. No. 470)}, vol.~2, pp.
  850--855, 1999.

\bibitem{Rajendran2018}
S.~{Rajendran}, W.~{Meert}, D.~{Giustiniano}, V.~{Lenders}, and S.~{Pollin},
  ``Deep learning models for wireless signal classification with distributed
  low-cost spectrum sensors,'' \emph{IEEE Transactions on Cognitive
  Communications and Networking}, vol.~4, no.~3, pp. 433--445, 2018.

\bibitem{Oshea2018}
T.~J. {O'Shea}, T.~{Roy}, and T.~C. {Clancy}, ``Over-the-air deep learning
  based radio signal classification,'' \emph{IEEE Journal of Selected Topics in
  Signal Processing}, vol.~12, no.~1, pp. 168--179, Feb 2018.

\bibitem{VGG}
K.~{Simonyan} and A.~{Zisserman}, ``Very deep convolutional networks for
  large-scale image recognition,'' \emph{arXiv e-prints}, p. arXiv:1409.1556,
  2014.

\bibitem{RN}
K.~{He}, X.~{Zhang}, S.~{Ren}, and J.~{Sun}, ``Deep residual learning for image
  recognition,'' \emph{2016 IEEE Conference on Computer Vision and Pattern
  Recognition (CVPR)}, pp. 770--778, 2016.

\bibitem{SRajendran2018}
S.~{Rajendran}, R.~{Calvo-Palomino}, M.~{Fuchs}, B.~{Van den Bergh},
  H.~{Cordobes}, D.~{Giustiniano}, S.~{Pollin}, and V.~{Lenders},
  ``Electrosense: Open and big spectrum data,'' \emph{IEEE Communications
  Magazin}, vol.~56, no.~1, pp. 210--217, 2018.

\bibitem{Xu2019}
T.~Xu and I.~Darwazeh, ``Deep learning for over-the-air non-orthogonal signal
  classification,'' \emph{arXiv:1911.06174}, 2019.

\bibitem{goodfellow2016deep}
I.~Goodfellow, Y.~Bengio, and A.~Courville, \emph{Deep Learning}.\hskip 1em
  plus 0.5em minus 0.4em\relax MIT Press, 2016.

\bibitem{Vincent2008}
P.~Vincent, H.~Larochelle, Y.~Bengio, and P.-A. Manzagol, ``Extracting and
  composing robust features with denoising autoencoders,'' \emph{Proceedings of
  the 25th International Conference on Machine Learning}, pp. 1096--1103, 2008.

\bibitem{chen2018}
M.-Y. Chen, T.-C. Huang, V.~Shu, C.-C. Chen, T.-C. Hsieh, and N.~Yen,
  ``Learning the chinese sentence representation with lstm autoencoder,''
  \emph{Proceedings of WWW '18: The Web Conference}, 2018.

\bibitem{ke2020}
Z.~Ke and H.~Vikalo, ``A graph auto-encoder for haplotype assembly and viral
  quasispecies reconstruction,'' \emph{Proceedings of The Thirty-Fourth AAAI
  Conference on Artificial Intelligence}, pp. 719--726, 2020.

\bibitem{Adam2014}
D.~Kingma and J.~Ba, ``Adam: A method for stochastic optimization,''
  \emph{International Conference on Learning Representations}, 12 2014.

\bibitem{Tomothy2016}
T.~O'Shea and N.~West, ``Radio machine learning dataset generation with gnu
  radio,'' \emph{Proceedings of the GNU Radio Conference}, vol.~1, no.~1, 2016.

\bibitem{OShea2016}
T.~J. {O'Shea}, J.~{Corgan}, and T.~C. {Clancy}, ``Unsupervised representation
  learning of structured radio communication signals,'' \emph{2016 First
  International Workshop on Sensing, Processing and Learning for Intelligent
  Machines (SPLINE)}, pp. 1--5, 2016.

\end{thebibliography}
\end{document}